\documentclass[prb,twocolumn,floatfix,superscriptaddress,noeprint]{revtex4-1}

\usepackage{graphicx,tabularx}
\usepackage{amsmath,amsfonts,amssymb}
\usepackage{bm,dsfont}
\usepackage[colorlinks=true, citecolor=blue, linkcolor=red, urlcolor=blue]{hyperref}
\usepackage[all]{hypcap}
\usepackage{CJK}
\usepackage{xcolor}
\usepackage[caption=false]{subfig}

\graphicspath{{figures/}}

\newcolumntype{C}{>{\centering\arraybackslash}X}
\newcolumntype{s}{>{\hsize=.2\hsize}X}

\begin{document}

\begin{CJK}{UTF8}{bsmi}
\title{Valley splitter and transverse valley focusing in twisted bilayer graphene}
\author{Christophe De Beule}
\email{c.de-beule@tu-braunschweig.de}
\affiliation{Institute for Mathematical Physics, TU Braunschweig, 38106 Braunschweig, Germany}
\author{Peter G. Silvestrov}
\email{p.silvestrov@tu-braunschweig.de}
\affiliation{Institute for Mathematical Physics, TU Braunschweig, 38106 Braunschweig, Germany}
\author{Ming-Hao Liu (劉明豪)}
\email{minghao.liu@phys.ncku.edu.tw}
\affiliation{Department of Physics, National Cheng Kung University, Tainan 70101, Taiwan}
\author{Patrik Recher}
\email{p.recher@tu-braunschweig.de}
\affiliation{Institute for Mathematical Physics, TU Braunschweig, 38106 Braunschweig, Germany}
\affiliation{Laboratory for Emerging Nanometrology, 38106 Braunschweig, Germany}
\date{\today}

\begin{abstract}
We study transport in twisted bilayer graphene and show that electrostatic barriers can act as valley splitters, where electrons from the $K$ ($K'$) valley are transmitted only to e.g.\ the top (bottom) layer, leading to valley-layer locked currents. We show that such a valley splitter is obtained when the barrier varies slowly on the moir\'e scale and induces a Lifshitz transition across the junction, i.e.\ a change in the Fermi surface topology. Furthermore, we show that for a given valley the reflected and transmitted current are transversely deflected, as time-reversal symmetry is effectively broken in each valley separately, resulting in valley-selective transverse focusing at zero magnetic field.
\end{abstract}

\maketitle
\end{CJK}

\section{Introduction} 

Twisted bilayer graphene (TBG) hosts a rich phenomenology ranging from its single-particle properties \cite{LopesDosSantos2007,Schmidt2008,Li2010,Bistritzer2010,Shallcross2010,Mele2010,SuarezMorell2010,TramblyDeLaissardiere2010,Mele2011,Luican2011,Brihuega2012,Moon2012,Yan2012,
Ohta2012,Moon2013,San-Jose2013,Schmidt2014,Kim2016,Cao2016,Nam2017,Huang2018,Rickhaus2018,Yuan2019} to the recent discoveries of correlated insulating phases \cite{Kim2017a,Cao2018a}, superconductivity \cite{Cao2018,Yankowitz2019}, nematicity \cite{Kerelsky2019,Choi2019}, and ferromagnetism \cite{Sharpe2019}. In TBG, two graphene layers are stacked at an angle that differs from Bernal ($AB$) or hexagonal ($AA$) stacking, giving rise to a moir\'e pattern of alternating stacking regions. For small twist angles, the moir\'e pattern varies slowly relative to the interatomic scale and the low-energy electronic properties are significantly modified \cite{LopesDosSantos2007,Bistritzer2010}. In this regime, the bands near charge neutrality become increasingly narrow with decreasing twist angles, accompanied by the reduction of the Fermi velocity \cite{Luican2011,Yin2015}, which vanishes at the so-called magic angle, and the emergence of low-energy Van Hove singularities (VHSs) \cite{Li2010,Brihuega2012,Ohta2012,Yan2012}.
\begin{figure}
\centering
\includegraphics[width=\linewidth]{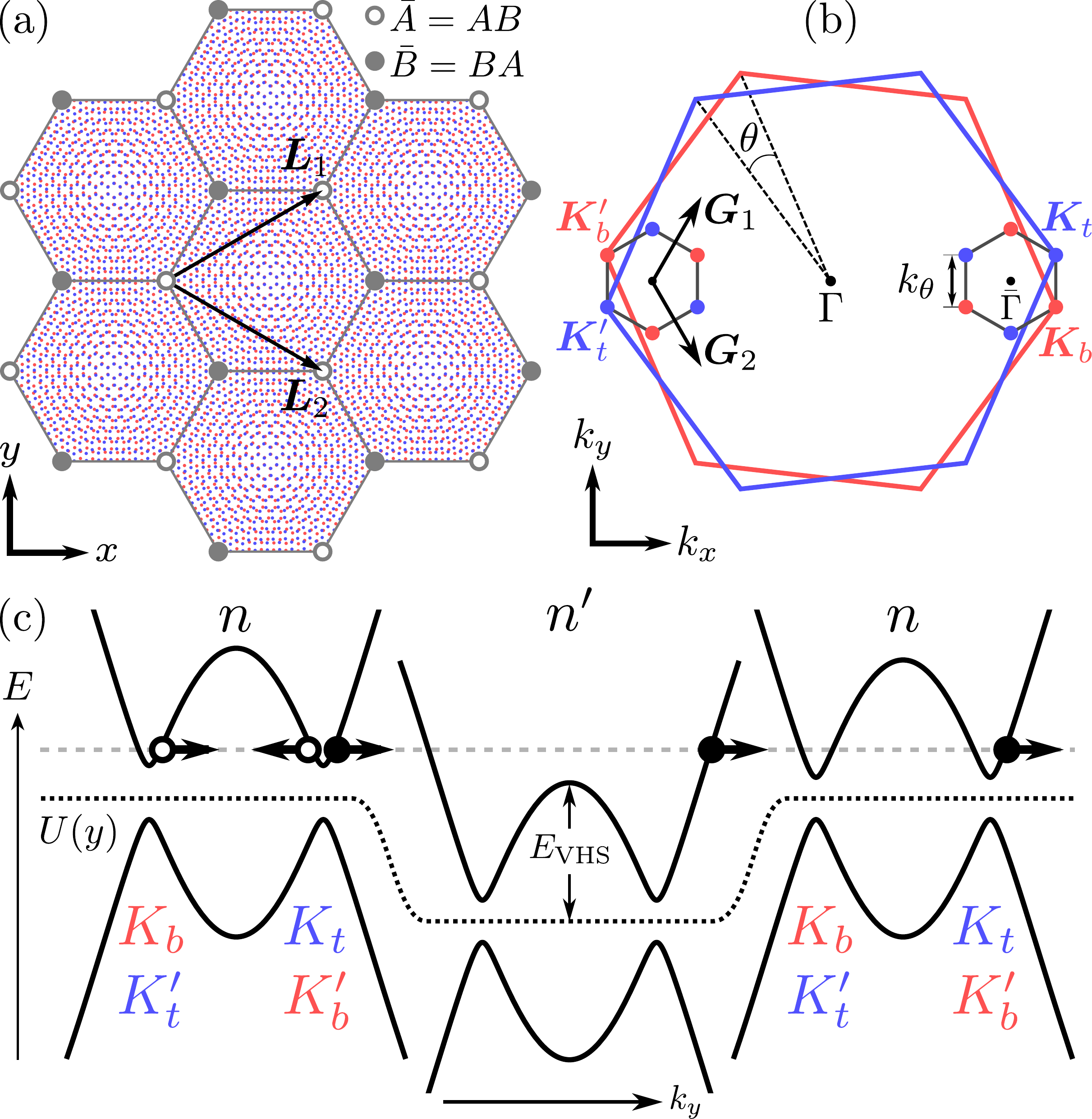} 
\caption{(a) TBG lattice for $\theta=3.9^\circ$ where blue (red) dots are atoms of the top (bottom) layer with the effective honeycomb lattice superimposed (gray dots). (b) Reciprocal space: large (small) hexagons are the rotated graphene (moir\'e) Brillouin zones and blue (red) dots are top (bottom) layer Dirac points. (c) Scattering at a potential barrier $U(y)$ (dotted line) in a wide ribbon along the zigzag direction of the effective lattice. Arrows indicate the propagation direction and the dashed line correspond to the Fermi energy. The potential induces a Lifshitz transition such that the transmitted current is valley-layer polarized as indicated.} \label{fig:fig1} 
\end{figure}

In this paper, we argue that TBG for intermediate twist angles ($2^\circ<\theta<10^\circ$), for which the layers are weakly coupled at low energies but strongly coupled near the VHS, is also an ideal platform for valleytronics \cite{Schaibley2016} where the two graphene layers provide a natural setting to separate the valleys. Contrary to valleytronics proposals in monolayer graphene which rely on specific edge termination or defects \cite{Rycerz2007,Akhmerov2008,Gunlycke2011}, we demonstrate that potential barriers in bulk TBG can generate valley-layer locked currents when the potential varies slowly on the moir\'e scale and when it induces a Lifshitz transtion across the junction. This corresponds to a change in the Fermi-surface topology, i.e.\ from two disconnected circles below the VHS to a closed curve above the VHS. Hence, our proposal should be robust against perturbations arising from lattice relaxation, displacement fields, or interactions, for example, as long as they do not remove the VHS. At intermediate twist angles, lattice relaxation does not change the band structure qualitatively \cite{Nam2017} and a displacement field perpendicular to the layers which induces an interlayer bias does not open a gap because the Dirac points of the two layers are separated in momentum space. Indeed, the existence of the VHS has been demonstrated experimentally for intermediate twist angles \cite{Li2010,Brihuega2012,Schmidt2014,Kim2016,Cao2016}. By putting two such \emph{valley splitters} in series, one can also design a valley valve \cite{Rycerz2007,Akhmerov2008}. Furthermore, as time-reversal symmetry is effectively broken within a single valley of TBG, the layer-locked current for a given valley attains a net transverse deflection, giving rise to transverse valley currents \cite{LopesDosSantos2012a}. The magnitude of deflection can be controlled by the slope of the potential and hence this effect, which we call \emph{transverse valley focusing}, can be thought of as an electronic counterpart of transverse magnetic focusing \cite{Taychatanapat2013,Berdyugin2019}.

The paper is organized as follows: In Section \ref{sec:theory}, we first introduce the TBG system and present the basic idea of the valley splitter using semiclassical arguments. We then solve the scattering problem with an effective two-band lattice model for TBG \cite{Yuan2018,Koshino2018,Po2018} that qualitatively reproduces the energy bands. We then discuss signatures of the valley splitter in the two-terminal conductance in Section \ref{sec:results} and demonstrate the transverse deflection of the valley-layer locked current with a multiterminal transport setup. Finally, we consider the effects of charge disorder on the scale of the moir\'e lattice. Further aspects of the models and transport properties are given in several appendices.

\section{Theory} \label{sec:theory}

\subsection{Twisted bilayer graphene}

Twisted bilayer graphene can be obtained by stacking two graphene layers in perfect registry and rotating the top (bottom) layer over an angle $\theta/2$ ($-\theta/2$), e.g.\ about a common hexagon center \cite{LopesDosSantos2007}. This results in a moir\'e pattern of alternating stacking regions with moir\'e lattice constant $L_m = a/[2\sin(\theta/2)]$ where $a\approx2.46$~\AA~is the graphene lattice constant, as shown in Fig.\ \ref{fig:fig1}(a). For small $\theta$, i.e.\ $L_m \gg a$, the low-energy physics becomes independent of twist center or commensuration due to approximate symmetries of the moir\'e pattern \cite{Bistritzer2010,LopesDosSantos2012a,Zou2018,Po2018}. It therefore suffices to only consider the resulting moir\'e lattice for a given twist angle in this limit. In Fig.\ \ref{fig:fig1}(a), the primitive moir\'e lattice vectors are given by $\bm L_{1,2} = L_m \left( \sqrt{3}/2, \pm 1/2 \right)$ and the corresponding reciprocal basis $\bm G_{1,2}$ is defined by $\bm G_i \cdot \bm L_j = 2\pi \delta_{ij}$. Since the moir\'e pattern varies slowly ($\sim a/\theta)$ relative to the interatomic scale for small twists, interlayer scattering between valleys is suppressed. The low-energy physics is then well described by a continuum model that treats the two valleys, which are related by time-reversal symmetry, independently \cite{LopesDosSantos2007,Bistritzer2010,LopesDosSantos2012a,Moon2013}. In momentum space, shown in Fig.\ \ref{fig:fig1}(b), the Dirac points of the rotated layers are given by $\bm K_l^{(\prime)} = R_{\pm \theta/2} \bm K^{(\prime)}$ for layer $l=t,b$, where $R_{\theta}$ is the standard rotation matrix, $\bm K^{(\prime)} = \tau |\bm K| \bm e_x$ are the unrotated Dirac points with $|\bm K| = 4\pi/(3a)$, and $\tau=\pm$ for the $K$ and $K'$ valley, respectively. We choose the moir\'e Brillouin zone (MBZ) such that $\bm K_l^{(\prime)}$ sit at the corners, and refer to these low-energy regions as minivalleys $\bar K / \bar K' = K_t/K_b$ or $K_b'/K_t'$ for the other valley, separated by $k_\theta = 2|\bm K|\sin(\theta/2)$ and where the bar indicates that these points lie in the MBZ.

\subsection{Valley splitter}

\begin{figure}
\centering
\includegraphics[width=\linewidth]{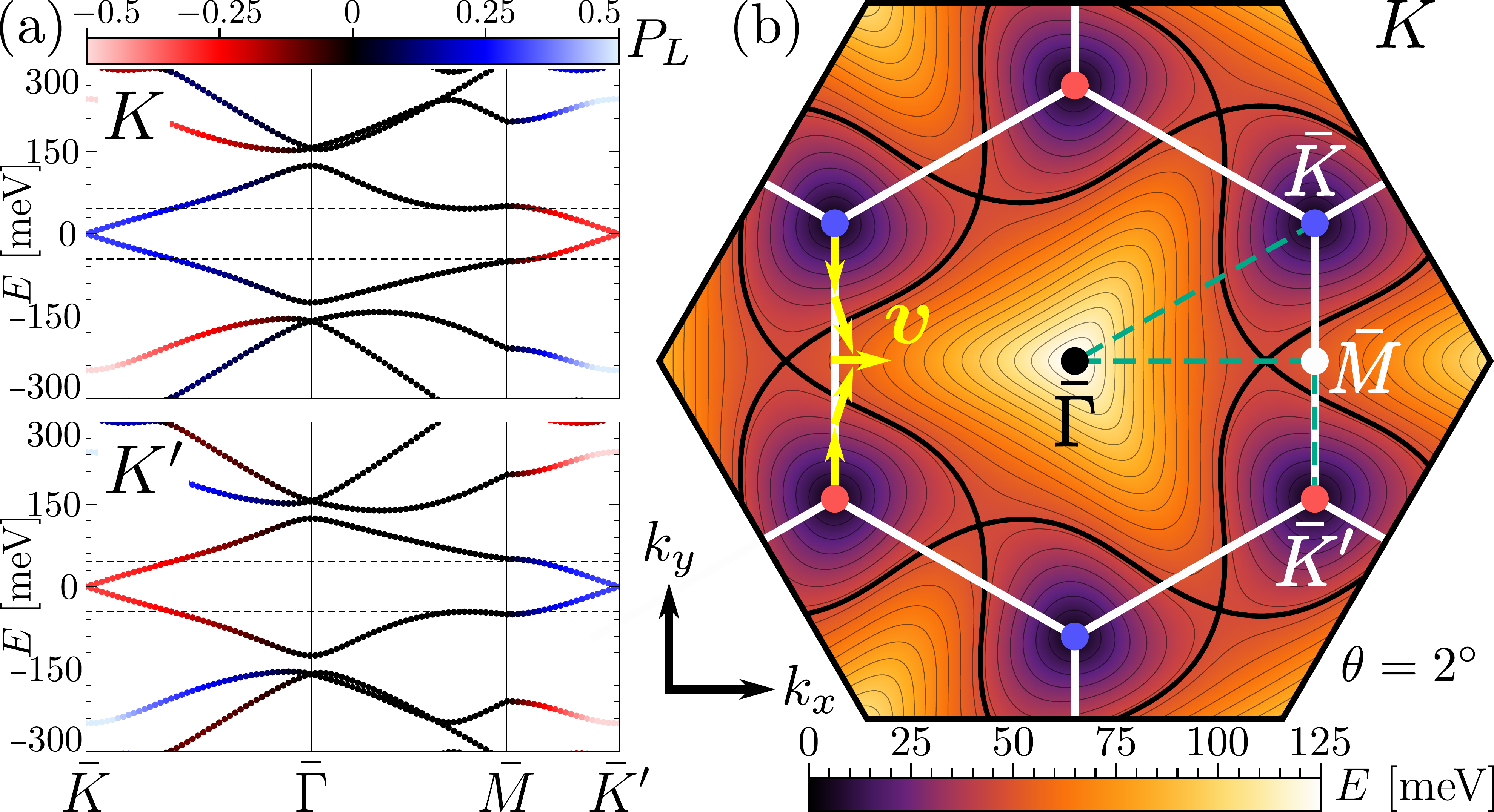}
\caption{Band structure for $\theta=2^\circ$. (a) Low-energy bands and layer polarization $P_L$ (see Eq.\ \eqref{eq:PL}) along the path shown in (b) as the green dashed line for the $K$ (top panel) and $K'$ (bottom panel) valley. The VHSs are indicated by horizontal dashed lines at $\pm E_{\textrm{VHS}} \approx \pm 46$~meV. (b) Lowest conduction band for $K$, where the thick contour corresponds to $E_{\textrm{VHS}}$, the inner white hexagon to the MBZ, and the group velocity $\bm v = \hbar^{-1} \nabla_{\bm k} E_{\bm k}$  near $\bar M$ is shown as the yellow arrows.} \label{fig:fig2} 
\end{figure}

In Fig.\ \ref{fig:fig1} (c), we schematically show a scattering process at a potential barrier in TBG oriented along the zigzag direction of the effective honeycomb lattice dual to the moir\'e lattice, given by the $y$-direction in Fig.\ \ref{fig:fig1}(a). Here, we consider the case where the Fermi energy $E_F$ is close to charge neutrality in the outer regions, while inside the barrier $E_F > E_{\bar M}$, where $E_{\bar M}$ is the energy of the lowest conduction band at the $\bar M$ point, which is defined in Fig.\ \ref{fig:fig2} where the low-energy bands calculated with the continuum model for $\theta=2^\circ$ are shown (see Appendix \ref{app:model} for the continuum model). As the potential $U(y)$ lies in the zigzag direction of the dual honeycomb lattice, inter-minivalley scattering is kinematically allowed. Now suppose there is a small bias voltage, such that electrons incident on the barrier arrive from the negative $y$-direction. When the potential varies slowly on the scale of $\lambda_F \propto E_F^{-1}$, the scattering process can be treated semiclassically. In this case, we expect that incident modes from $K_b$ ($K_t'$) are always reflected back to $K_t$ ($K_b'$) as they pass through a local extremum, i.e.\ a classical turning point. Moreover, because the minivalleys correspond to the layers at low energies, the transmitted current runs mostly in the bottom (top) layer for the $K$ ($K'$) valley. The total transmitted current therefore consists of a $K$-polarized current in the top layer and a $K'$-polarized current in the bottom layer, where the valley polarization is reversed by reversing the bias voltage or the sign of the potential. Furthermore, we find that incident modes at $K_b$ ($K_t'$) that are reflected to $K_t$ ($K_b'$) are transversely deflected, which can be understood from the transverse velocity, as illustrated in Fig.\ \ref{fig:fig2}(b). This gives rise to transverse valley currents and valley-dependent deflection and is a consequence of time-reversal symmetry breaking in a single valley of TBG.

The range of twist angles for which the valley splitter can be realized is determined by the carrier density $n_{\bar M} \approx k_\theta^2/(2\pi)$ required for the Lifshitz transition across the junction, and whether states near the Dirac points are sufficiently localized in their respective layer. The former condition gives the upper limit as $\theta \approx 3^\circ \sqrt{n_{\bar M}}$ for $n_{\bar M}$ in units $10^{13}$ cm$^{-2}$. For the latter condition, we calculate the layer polarization of the Bloch states with the continuum model. The Bloch states can be written as $\Psi_{m\bm k}(\bm r) = ( 1/\sqrt{V} ) \sum_{\bm G} \left( \psi_{t,m\bm k - \bm G}, \, \psi_{b,m\bm k - \bm G} \right)^t e^{i \left( \bm k - \bm G \right) \cdot \bm r}$, where $m$ is a band index, $\bm k$ is the crystal momentum constrained to the first MBZ, $V$ is the sample area, $\bm G = n_1 \bm G_1 + n_2 \bm G_2$ ($n_{1,2} \in \mathbb Z$) is a reciprocal lattice vector of the moir\'e lattice, and $\psi_l$ are two-component spinors containing the amplitudes on the $A/B$ sublattices of layer $l=t,b$.
The layer polarization is then given by
\begin{equation} \label{eq:PL}
P_{L,m}(\bm k) = \sum_{\bm G} \left( | \psi_{t,m\bm k - \bm G} |^2 - | \psi_{b,m\bm k - \bm G} |^2 \right),
\end{equation}
and is shown in Fig.\ \ref{fig:fig2}(a) for $\theta = 2^\circ$. At the Dirac points, $P_L(\bar K / \bar K') = \pm 0.32$ for the $K$ valley, so that $(1\pm P_L)/2=66\%$ of the density remains localized in its layer, which increases up to $82\%$ for $\theta = 3^\circ$. Hence, the valley splitter regime roughly corresponds to the range $2^\circ<\theta< 10^\circ$, where the lower (upper) limit is determined by $P_L$ ($n_{\bar M}$). Devices with twist angles in this range have already been realized in several experiments \cite{Li2010,Brihuega2012,Schmidt2014,Kim2016,Cao2016}.
\begin{table}[t!]
\centering
	\begin{tabularx}{.6\linewidth}{s | C}
	\hline \hline
	& $\theta=2^\circ$ \\
	\hline
	$L_m$ & $7.05$~nm \\
	$t_1$ & $33.24$~meV \\
	$t_2'/t_1$ & $0.069$ \\
	$t_4/t_1$ & $-0.089+0.209i$ \\
	$t_5/t_1$ & $+0.104+0.062i$ \\
	$t_6/t_1$ & $+0.017+0.022i$ \\
	\hline \hline
	\end{tabularx}
\caption{Five largest hopping parameters of the lattice model \eqref{eq:latticemodel} and moir\'e lattice constant $L_m$ for $\theta=2^\circ$. The hoppings are defined in Fig.\ \ref{fig:sfig2}(b) with $t_2\approx it_2'$.}
\label{tab:hopping}
\end{table}

\subsection{Effective lattice model} 

We study electronic transport with a two-band lattice model for each valley separately, which we obtain with the Wannier method of Koshino \emph{et al.}\ \cite{Koshino2018} and which describes the two (spin degenerate) bands near charge neutrality \cite{Yuan2018}. The lattice model is defined on an effective honeycomb lattice with sublattices $\bar A/\bar B$ corresponding to $AB/BA$ stacking regions of TBG (Fig.\ \ref{fig:fig1}(a)). The effective Hamiltonian for the two low-energy bands can be written as
\begin{equation} \label{eq:latticemodel}
\hat H_{\textrm{eff}} = \sum_{nn'} \sum_{\sigma\sigma'}  t_{nn'}^{\sigma \sigma'} \hat c_{n\sigma}^\dag \hat c_{n'\sigma'},
\end{equation}
where $n,m$ run over the sites of the effective honeycomb lattice with $\sigma=\bar A,\bar B$. We also defined fermion operators $\hat c_{n\sigma}$ ($\hat c_{n\sigma}^\dag$) which destroy (create) particles in the Wannier state $\left| n , \sigma \right>$ and hopping matrices $t_{nn'}^{\sigma \sigma'}$. See Appendix \ref{app:2band} for more details concerning the two-band model.

We then consider transport in the zigzag direction of the effective honeycomb lattice. For a clean system without edges, we can Fourier transform \eqref{eq:latticemodel} along the transverse direction ($x$ direction in Fig.\ \ref{fig:fig1}(a)),
\begin{equation}
\hat c_{n\sigma} = \frac{1}{\sqrt{N_x}} \sum_k e^{ikL_x n_x} \hat c_{kn_y\sigma},
\end{equation}
where $n=(n_x,n_y)$ and $L_x = \sqrt{3} L_m$ is the lattice constant of an armchair ribbon. This results in an effective one-dimensional (1D) chain with two orbitals per cell and hopping matrices $\tilde t_{n_yn_y'}(k)$ ($\tilde t_{n_yn_y'}(-k)^*$ for the other valley) with $k$ the transverse momentum. We obtain
\begin{equation} \label{eq:hamL}
\hat H_{\textrm{1D}}(k) = \sum_{n,n'} \sum_{\sigma \sigma'} \left[ \delta_{nn'} \delta_{\sigma\sigma'} U_n + \tilde t_{nn'}^{\sigma\sigma'}(k) \right] \hat c_{n'\sigma'}^\dag \hat c_{n\sigma}
\end{equation}
where we dropped the subscript $y$ for convenience and $U_n$ is the onsite energy. The hopping matrices are
\begin{equation}
\tilde t_{\textrm{intracell}} = \begin{pmatrix}
2t_2' \sin k & z \\
z^* & 2t_2' \sin k \end{pmatrix},
\end{equation}
with $z=t_1 e^{i\frac{k}{2}} + t_4^* e^{-i\frac{k}{2}} + t_6^* e^{i\frac{3k}{2}}$ and
\begin{align}
\tilde t_{\textrm{inter1}} & =
\begin{pmatrix}
0 & t_1 + t_5^* e^{ik} \\
t_1 + t_5 e^{-ik} & 0 \end{pmatrix}, \\
\tilde t_{\textrm{inter2}} & = 
\begin{pmatrix} 
0 & t_4^* e^{i\frac{k}{2}} + t_5^* e^{-i\frac{k}{2}} \\
t_4 e^{-i\frac{k}{2}} + t_5 e^{i\frac{k}{2}} & 0 \end{pmatrix}, \\
\tilde t_{\textrm{inter3}} & = 
\begin{pmatrix}
-2t_2' \sin \frac{k}{2} & t_5^* \\ t_5 & -2t_2' \sin \frac{k}{2} \end{pmatrix}, \\
\tilde t_{\textrm{inter4}} & = \begin{pmatrix} 0 & t_6^* e^{-i\frac{k}{2}} \\ t_6 e^{i\frac{k}{2}}  & 0 \end{pmatrix},
\end{align}
where the index denotes $n$th-nearest intercell hopping and we only consider the five largest hopping amplitudes of the effective lattice model.

The hopping amplitudes are calculated with the method of Ref.\ \onlinecite{Koshino2018} and are given in Table \ref{tab:hopping} where we have chosen the nearest-neighbor hopping amplitude $t_1$ real and positive. We model the potential barrier as $U_n  = ( U_0 / 2) \left( f \left[ (n - n_l)/2d \right] - f \left[ (n - n_r)/2d \right] \right)$, where $f(y) = \tanh(y)$, $d$ determines the variation of the potential in units $L_m$, and the length of the barrier is given by $L =(L_m/2) \left( n_r - n_l \right)$. In the limit $d\rightarrow0$, the potential becomes sharp on the moir\'e scale. Other functional forms for the potential barrier are also possible \cite{Rycerz2007}. However, we believe the results will not change qualitatively as the important feature is the length scale over which the potential varies. This length scale is fixed for a specific device as it depends on the details of the gate, while the barrier height $U_0$ can be tuned by a gate voltage. Typical values for $d$ range between a few nm to a few tens of nm for encapsulated graphene samples, depending on the distance between graphene and the gate electrode. By intentionally using thicker dielectrics, it is possible to make $d$ even larger. When the graphene sample is suspended, the smoothness can be as large as several hundreds of nm, see for example Ref.\ \onlinecite{Rickhaus2013}. 

Henceforth, we focus on $\theta=2^\circ$ for which $L_m \approx 7.05$~nm. This is motivated by the fact that transport experiments at this twist angle have already been performed, where the Lifshitz transition was observed as a reversal of the sign of the Hall conductance \cite{Kim2016,Cao2016}. At larger twist angles, the valley splitter is expected to work better because the minivalleys become more localized in their respective layers. However, the required density to reach the VHS also increases, making the experimental realization more demanding. We also focus on valley $K$, as defined in Fig.\ \ref{fig:fig1}. Results for $K'$ can be obtained from  $K$, as the valleys are decoupled for small twists and related by time-reversal symmetry.
\begin{figure}
\centering
\includegraphics[width=\linewidth]{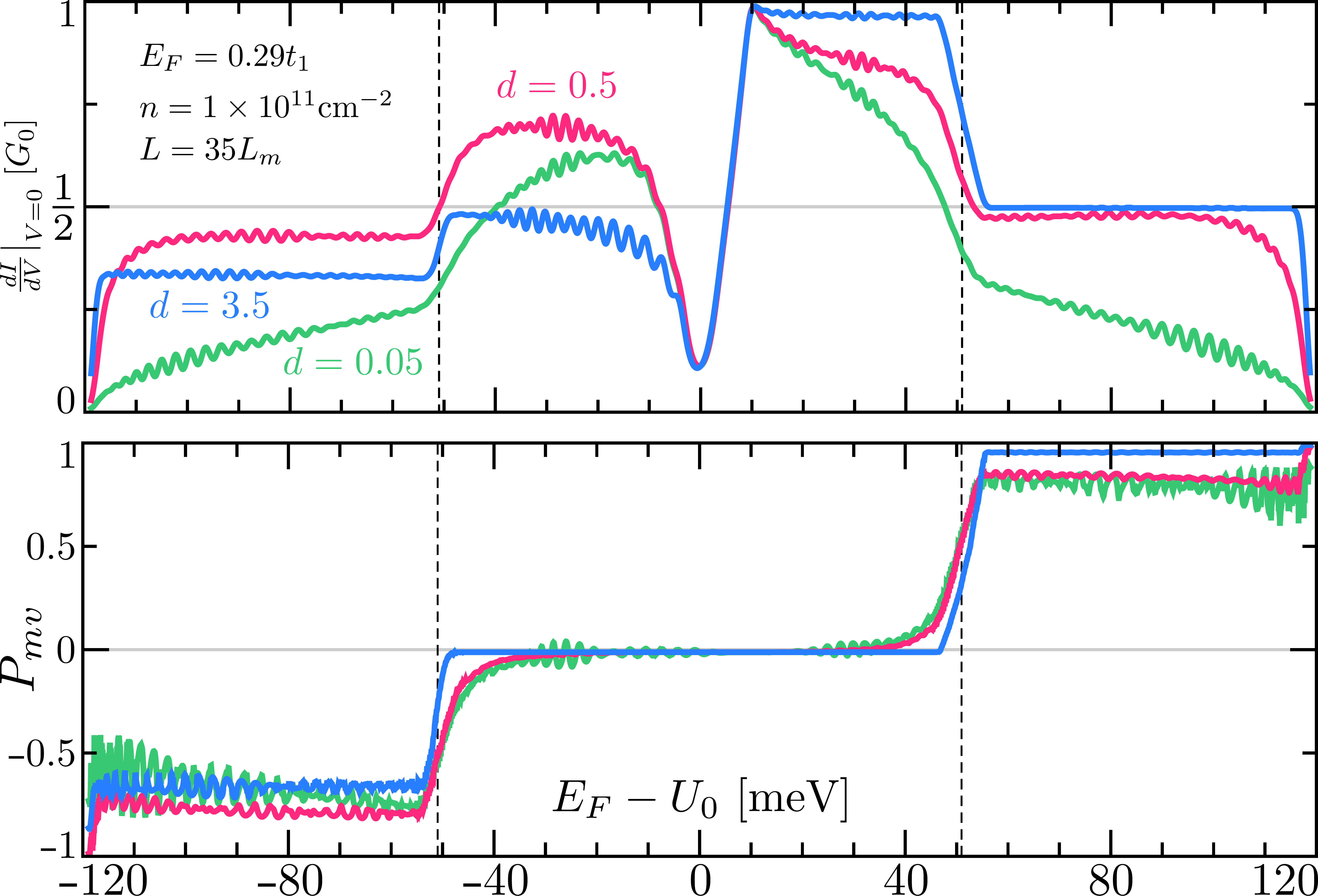}
\caption{(top) Normalized conductance for states at the $K$ valley as a function of the Fermi energy in the barrier $E_F-U_0$ for fast ($d<1$) and slowly ($d>1$) varying potentials relative to the moir\'e scale. Interference oscillations are suppressed by Gaussian averaging (raw data in Appendix \ref{app:2terminal})
and the vertical dashed lines correspond to $\pm E_{\bar M}$. (bottom) Minivalley polarization of the transmitted modes as defined in Eq.\ \eqref{eq:mv}. Here, the Fermi wave length $\lambda_F \approx 3.6L_m$.} \label{fig:fig3}
\end{figure}

\section{Results} \label{sec:results}

\subsection{Two-terminal setup}

The one-dimensional scattering problem for the two-terminal setup was solved with \textsc{Kwant} \cite{Groth2014} and the resulting zero-bias differential conductance is shown in the top panel of Fig.\ \ref{fig:fig3} as a function of the Fermi energy in the barrier $E_F-U_0$. When the barrier varies slowly on the moir\'e scale ($d>1$) and $E_F-U_0>E_{\bar M}$, the conductance reaches a plateau of $G_0/2$, where $G_0$ is the ballistic conductance per valley. This case corresponds to the schematic in Fig.\ \ref{fig:fig1}(c). We find that the barrier is transparent for one minivalley and opaque for the other, as is corroborated by the minivalley polarization
\begin{equation} \label{eq:mv}
P_{mv} = \frac{\sum_{k_x} \left( T_t - T_b \right)}{\sum_{k_x} \left( T_t + T_b \right)}, \qquad T_l = \sum_{l'=t,b} T_{l \leftarrow l'},
\end{equation}
where $T_{l \leftarrow l'}$ is the transmission function for scattering from minivalley $l'$ to $l$, and which is shown in the bottom panel of Fig.\ \ref{fig:fig3}. We identify these transmission functions from the longitudinal momentum of incoming and transmitted modes. When the momentum is positive (negative), we assign it to $K_t$ ($K_b$).

The beating pattern in the conductance is due to a mismatching resonance condition for modes at $\pm k_x$ as time-reversal symmetry is broken within a single valley of TBG. In general, interference oscillations are more prominent in the bipolar regime ($E_F-U_0<0$) due to reflections at $n$-$p$ junctions on both sides of the potential barrier \cite{Pereira2006a,Silvestrov2007}. For the same reason, the minivalley filter does not work perfectly for $E_F-U_0<-E_{\bar M}$ and in this case its efficiency decreases for the slowly-varying potential barrier ($d=3.5$, blue line). This can be understood as follows. When the barrier is smooth on the scale of $\lambda_F$, transmission through the {$n$-$p$} junction is suppressed away from normal incidence \cite{Cheianov2006}. Hence, the filter in the bipolar regime works better for barriers with intermediate smoothness ($d \approx1$, red line). Since a smooth barrier in the filter regime only transmits electrons from one minivalley for a given valley, one can try to design a valley valve \cite{Rycerz2007,Akhmerov2008} by placing two such barriers in series. When the barriers are gated similarly, one minivalley can go through. For opposite polarities, however, electrons that pass the first barrier are reflected and change minivalley in the second barrier, at which point they can pass the first barrier from the other side. More details of the minivalley filter and valley valve are given in Appendix \ref{app:2terminal}.

Note that so far, we assumed that the barrier is oriented along the zigzag direction of the dual honeycomb lattice, see Fig.\ \ref{fig:fig1}(a). In principle this could be achieved by visualizing the moir\'e pattern with scanning tunneling microscopy \cite{Li2010} and rotating the sample accordingly. Nevertheless, if the orientation differs by an angle $\chi$ relative to the zigzag direction, there is always a finite part of electrons at the Fermi surface demonstrating the same physics provided that $\tan \chi < 2k_F/k_\theta=\sqrt{2\pi n}/k_\theta$.
\begin{figure}
\centering
\includegraphics[width=\linewidth]{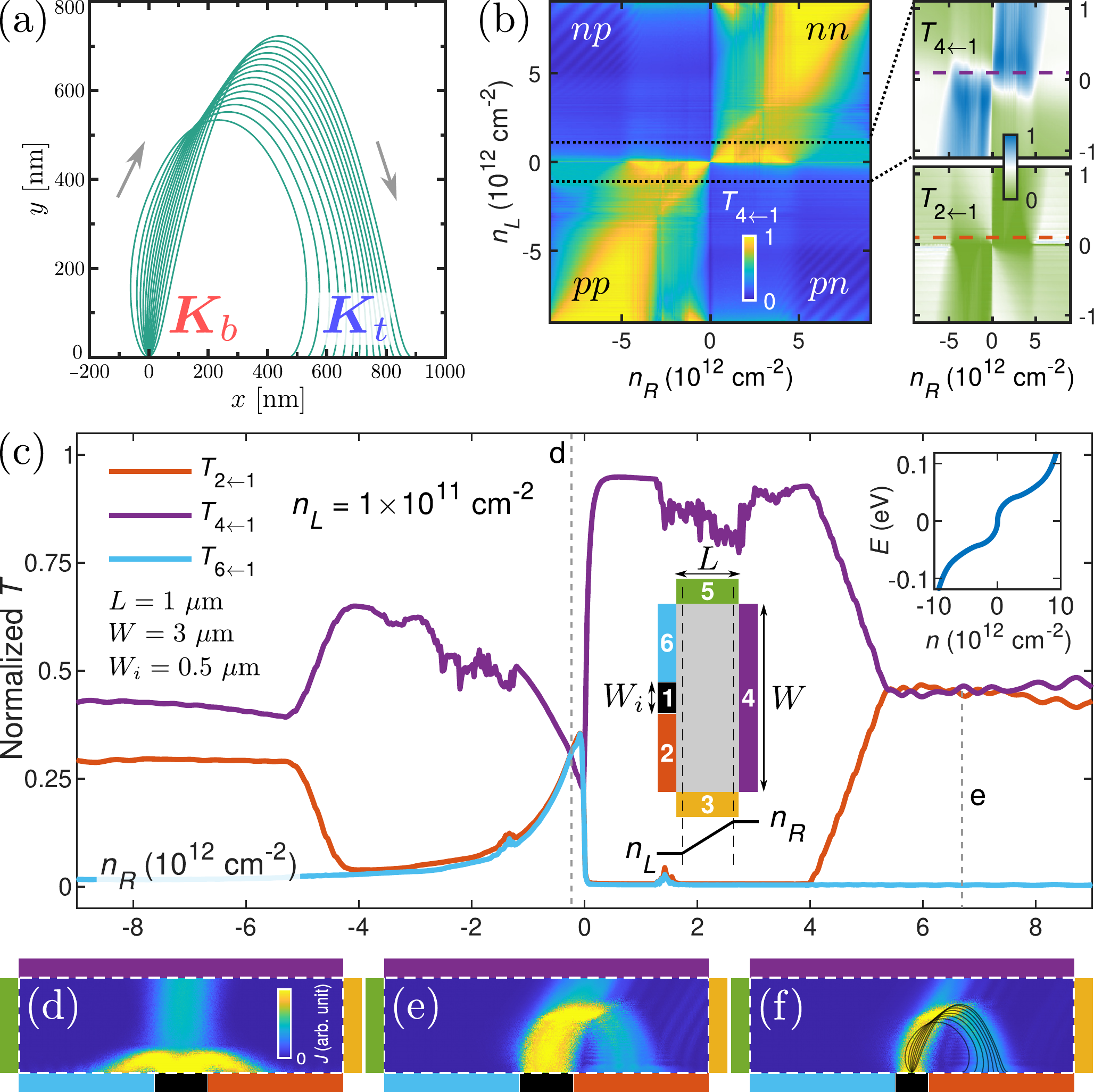}
\caption{(a) Classical trajectories for $K_b$ modes injected at $\bm r=(0,0)$ for $n_L (n_R) = 0.1 (6.69) \times 10^{12}$ cm$^{-2}$. Arrows indicate direction of motion. (b) $T_{4\leftarrow1}$ for the setup shown in (c) where the side panel zooms near neutrality and also shows $T_{2\leftarrow1}$. (c) Transmission versus $n_R$ for $n_L =10^{11}$ cm$^{-2}$ as indicated by the dashed line in (b). Corner inset shows the relation between energy and density. (d)--(e) Magnitude of the current density with $n_R$ given by the marked dashed lines in (c), respectively. (f) Same as (e) except for a smaller injector lead ($W_i = 0.3$~$\mu$m) and with (a) superimposed.}
\label{fig:fig4} 
\end{figure}

\subsection{Multiterminal setup}

To demonstrate the transverse deflection of the reflected current, we consider a multiterminal setup of length $L=1$~$\mu$m and width $W=3$~$\mu$m, as shown in the center inset of Fig.\ \ref{fig:fig4}(c), for a smooth potential step modeled by a linear carrier density profile $n(y) = n_L + \left( n_R - n_L \right) y / y_0$  for $0 < y < y_0$ and constant otherwise, where we take $y_0=0.9$ $\mu~\textrm{m} \approx 128 L_m$. The corresponding on-site energy profile follows from the energy-density relation shown in the upper-right inset of Fig.\ \ref{fig:fig4}(c). The multiterminal transport problem was solved with the Green's function method \cite{Datta1995} using our own code, see for example Refs.\ \onlinecite{Rickhaus2015a,Liu2017}. For clarity, all transmission functions are normalized by the number of incoming modes in lead $1$. In Fig.\ \ref{fig:fig4}(b), we show the transmission function $T_{4\leftarrow1}$ from lead $1$ to lead $4$ as a function of the density $n_R$ and $n_L$ on the right and left side of the potential step, respectively. The upper part of the side panel of Fig.\ \ref{fig:fig4}(b) zooms in near charge neutrality, and the lower part shows $T_{2\leftarrow 1}$ for comparison. When $n_L$ is near charge neutrality and $n_R$ corresponds to $E>E_{\bar M}$, we have $T_{4\leftarrow 1} = T_{2\leftarrow 1} = 1/2$ as before, which is better seen in Fig.\ \ref{fig:fig4}(c) where we show the transmission probability for $n_L=10^{11}$ cm$^{-2}$, corresponding to the dashed lines in the insets of Fig.\ \ref{fig:fig4}(b). We see that the plateau at $1/2$ for $n_R>5\times 10^{12}$ cm$^{-2}$ is imperfect, as few modes escape through other leads. Nevertheless, almost all reflected modes end up in lead 2, whereas $T_{6\leftarrow 1}$ is negligible. The transverse deflection is best visualized by the current density (see for example Refs.\ \onlinecite{Rickhaus2015a,Liu2017,Rickhaus2015} for technical details). The magnitude of the current density is shown in Fig.\ \ref{fig:fig4}(d) for $n_L=-n_R=10^{11}$ cm$^{-2}$ and shows a Klein-collimated electron beam \cite{Liu2017}, similar as in graphene $n$-$p$ junctions, due to the smooth junction \cite{Cheianov2006}. The current density in the valley splitter regime is shown in Figs.\ \ref{fig:fig4}(e) and (f), where in (f) the width of the injector lead is slightly reduced and the classical trajectories of the reflected current are superimposed. More exemplary current density images are given in Appendix \ref{app:6terminal}. The magnitude of the deflection can be tuned by changing the slope of the potential which controls the classical turning point. Importantly, the deflection is in the opposite direction for the other valley, which is equivalent to interchanging lead $2$ and lead $6$, as the valleys are related by time-reversal symmetry.

\subsection{Charge disorder on the moir\'e scale}

\begin{figure}
\includegraphics[width=\linewidth]{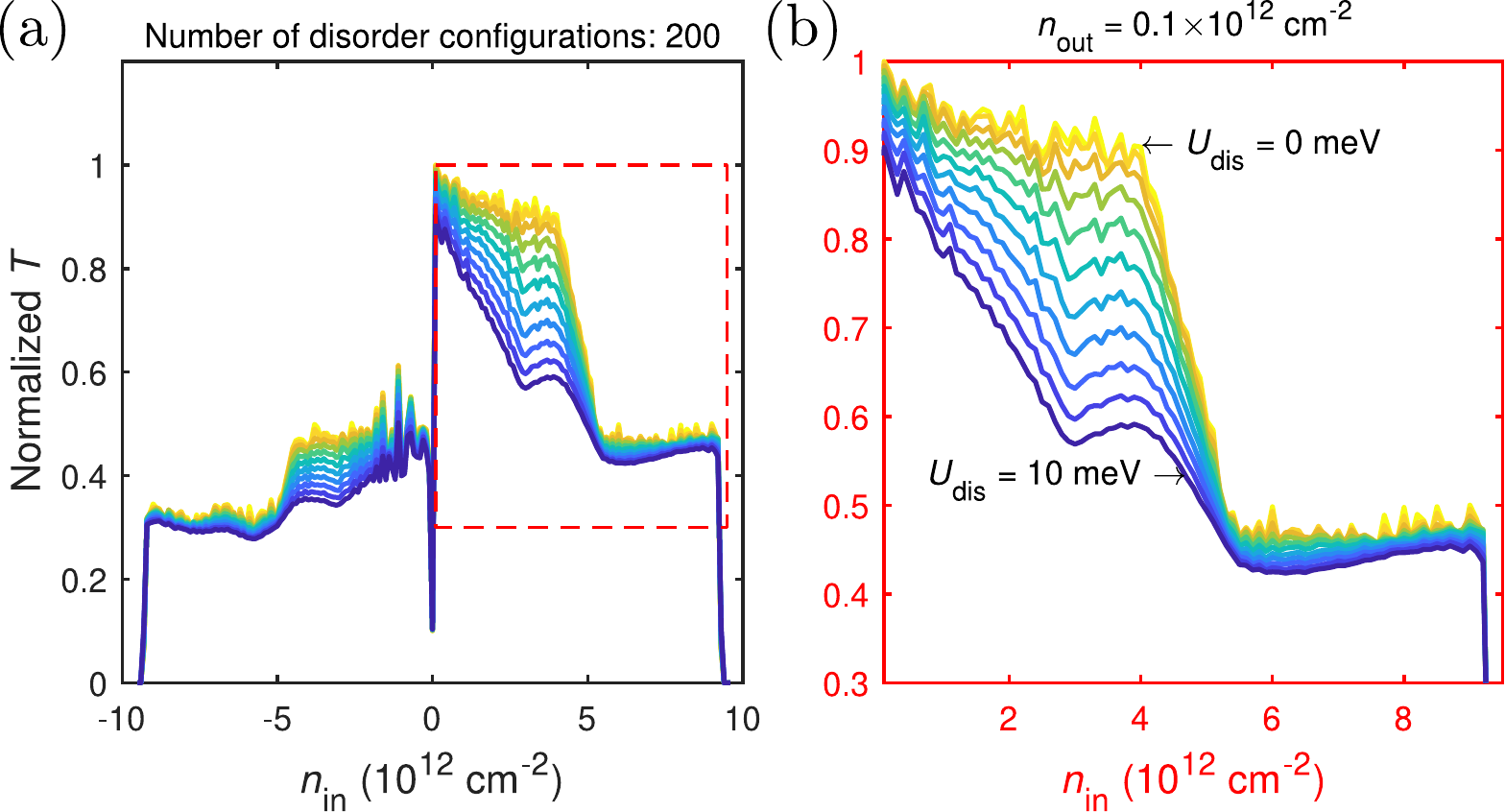}
\caption{(a) Transmission versus density in the barrier for the two-terminal setup [top panel of Fig.\ \ref{fig:fig3}] for smoothness $d=3.5$ but for a sample with finite width $W = 2$~$\mu$m and with random onsite energies according to a uniform distribution in $[-U_{\textrm{dis}},U_{\textrm{dis}}]$. The result is averaged over $200$ disorder configurations. The top yellow curve corresponds to no disorder, which increases to $U_{\textrm{dis}}=10$~meV for the bottom blue curve. (b) Close up of near  the unipolar minivalley filter regime marked in the left panel by the dashed box.}
\label{fig:figDisorder}
\end{figure}

Disorder on the moir\'e scale due to inhomogeneities in the charge density or twist angle mixes the minivalleys $K_t$ and $K_b$ (or $K_t'$ and $K_b'$) which reduces the efficiency of the valley splitter. However, as long as the disorder scale is below the VHS energy, the valley splitter is robust, as it only relies on a change in the Fermi-surface topology across the junction. In Fig.\ \ref{fig:figDisorder}, we show the transmission versus the density inside the barrier for the same setup as shown in the top panel of Fig.\ \ref{fig:fig3} for a barrier with smoothness $d=3.5$ but for a sample with finite width $W = 2$~$\mu$m and with random onsite energies. The onsite energies vary according to a uniform distribution between $[-U_{\textrm{dis}},U_{\textrm{dis}}]$ and we show the average transmission over $200$ different such disorder configurations. Note that the plateau in the transmission in the minivalley filter regime (in both the unipolar and bipolar case) is  remarkably stable compared to the transmission away from this regime, which quickly decays. This is a consequence of the robustness of the valley splitter, which only requires a Lifshitz transition across the junction. Hence, the valley splitter is unaffected as long as the disorder strength is well below the VHS energy.

\section{Conclusion} 

We have shown that electrostatic barriers in twisted bilayer graphene for intermediate twist angles give rise to valley-layer locked currents when the electrostatic potential induces a Lifshitz transition, i.e.\ a change in the Fermi-surface topology, across the junction. Such currents are characterized by a conductance plateau in each layer given by one fourth of the total ballistic conductance. Moreover, we have also demonstrated that the current is transversely focused in opposite directions for opposite valleys, which could be understood as time-reversal symmetry is effectively broken within a single valley of twisted bilayer graphene. This could be observed by injecting electrons from a narrow contact and measuring the response at nearby contacts. Our paper proposes a new means of manipulating the valley degree of freedom in graphene systems.

\begin{acknowledgments}
We thank R. J. Haug, A. Schuray, and F. Dominguez for helpful discussions, and K. Richter for hosting M.-H.L.\ for a scientific visit during which part of this work was performed. P.G.S. and P.R. thank the Deutsche Forschungsgemeinschaft (DFG, German Research Foundation) for funding through Grant No. RE 2978/8-1 and within the framework of Germany's Excellence Strategy---EXC-2123 QuantumFrontiers---390837967. M.-H.L. was supported by the Taiwan Ministry of Science and Technology (Grants No. 107-2112-M-006-004-MY3 and No. 109-2112-M-006-020-MY3).
\end{acknowledgments}

\appendix

\section{CONTINUUM MODEL} \label{app:model}

For small twist angles, the low-energy physics of twisted bilayer graphene (TBG) becomes independent of the geometric details of the lattice, and is well described by a continuum model \cite{Bistritzer2010,LopesDosSantos2012a,Moon2013}. In this model, the two valleys are treated separately as intervalley scattering is suppressed because the moir\'e pattern varies slowly on the atomic scale. Furthermore, only the dominant terms in the interlayer tunneling are retained. In real space, the continuum model for a single valley can be written as
\begin{equation} \label{eq:cont}
\hat H = \int d^2 r \, \hat \psi^\dag(\bm r) \begin{pmatrix} \hat H_t & T(\bm r) \\ T^\dag(\bm r) & \hat H_b \end{pmatrix} \hat \psi(\bm r),
\end{equation}
with $\hat H_t$ ($\hat H_b$) the Dirac Hamiltonian of the top (bottom) layer, respectively, and $T$ the interlayer coupling:
\begin{align}
\hat H_l & = \hbar v_F \bm \sigma \cdot \left[ R_{\mp\theta/2} \left( -i \nabla - \bm K_l \right) \right], \\
T(\bm r) & = T_0 + T_1 e^{i \bm G_1 \cdot \bm r} + T_2 e^{i \bm G_2 \cdot \bm r},  \label{eq:coupling}
\end{align}
where $v_F$ is the Fermi velocity of graphene, $\bm K_{t,b} = k_\theta (\sqrt{3}/2,\pm1/2)$, $\bm \sigma = (\sigma_x, \sigma_y)$ the Pauli matrices, and $R_\theta$ corresponds to a counterclokwise rotation over $\theta$. Here, we also have $\bm G_{1,2} = \sqrt{3} k_\theta \left( \pm 1/2, \sqrt{3}/2 \right)$ (note that $\bm G_2$ is defined with an opposite sign in the main text) for $\bm K = K \bm e_x$, which is a reciprocal basis of the moir\'e pattern and $k_\theta=2K \sin ( \theta/2 ) = 4\pi/(3L_m)$ with $K = 4\pi/(3a)$ where $a\approx0.246$~nm is the graphene lattice constant. The interlayer coupling matrices $T_n$ ($n=0,1,2$) are given by
\begin{equation}
T_ n = \begin{pmatrix} u_{AA} & u_{AB} \omega^n \\ \omega^{-n} u_{AB} & u_{AA} \end{pmatrix},
\end{equation}
with $\omega=e^{i2\pi/3}$ and $u_{AA}/u_{AB}\approx0.8$ accounts for the corrugation of the twisted bilayer \cite{Mele2011,Nam2017}. 

The Hamiltonian is then diagonalized by plane-wave expansion. To this end, we consider a general Bloch wave,
\begin{equation} \label{eq:bloch}
| \Psi_{\bm k} \rangle = \sum_{\bm G} \begin{pmatrix} \psi_{t,\bm k-\bm G} \\ \psi_{b,\bm k-\bm G} \end{pmatrix} | \bm k - \bm G \rangle,
\end{equation}
where $\bm G = n_1 \bm G_1 + n_2 \bm G_2$ with $n_{1,2}$ integers and $\psi_l$ are two-spinors containing amplitudes of $A_l$ and $B_l$ sublattices. In momentum space, \eqref{eq:cont} then becomes
\begin{equation} \label{eq:blochH}
\begin{aligned}
& \hat H = \sum_{\bm k,\bm G} \Bigg\{ \sum_{l=t,b} \hat c_{l,\bm k-\bm G}^\dag H_l(\bm k - \bm G) \hat c_{l,\bm k-\bm G} + \\
& \sum_{\bm G'} \left[ \sum_{n=0}^2 \delta_{\bm G',\bm G+\bm G_n} \hat c_{t,\bm k-\bm G}^\dag \, T_n \, \hat c_{b,\bm k-\bm G'} + \textrm{h.c.} \right] \Bigg\},
\end{aligned}
\end{equation}
where $\hat c_{l,\bm k-\bm G}^\dag$ ($\hat c_{l,\bm k-\bm G}$) are two-component fermion operators that create (destroy) a Dirac particle in a single valley in layer $l$ with momentum $\bm k - \bm G - \bm K_l$, $H_l(\bm q) = e^{-i\bm q \cdot \bm r} \hat H_l e^{i\bm q \cdot \bm r}$, and $\bm G_0 = \bm 0$. Here, $\bm k$ is constrained to the moir\'e Brillouin zone (MBZ) and defined with respect to the $\bar\Gamma$ point. Diagonalization of \eqref{eq:blochH} requires solving an infinite system of equations. In practice, to achieve convergence of the low-energy bands in the first MBZ, we only include $\bm G$ for which $|\bm K_l + \bm G| < 6 k_\theta$. In this way, we obtain the energy bands $E_{m\bm k}$, shown in Fig.\ \ref{fig:fig2} and Fig.\ \ref{fig:sfig3}(b) and the corresponding Bloch states $| \Psi_{m\bm k} \rangle$ with $m$ the band index, where we used the parameters of Ref.\ \onlinecite{Koshino2018}: $a=0.246$~nm, $\hbar v_F = 525.3$~meV nm, $u_{AA} = 79.7$~meV, and $u_{AB} = 97.5$~meV.

\begin{figure}
\centering
\includegraphics[width=.75\linewidth]{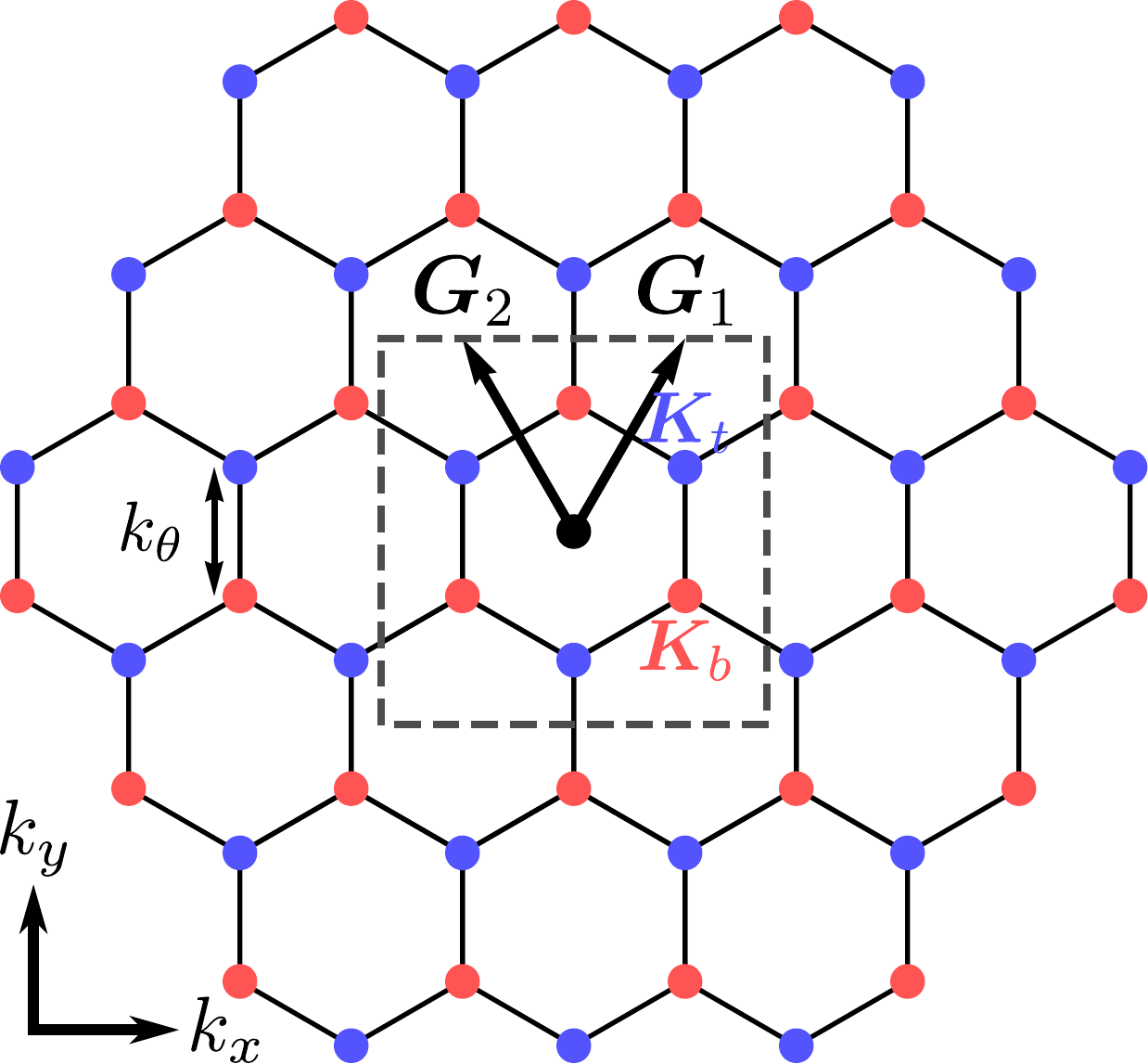}
\caption{Reciprocal space of the moir\'e lattice in the extended zone scheme. Blue (red) dots are Dirac points of the top (bottom) layer for one valley. In the continuum model, only scattering between neighboring Dirac cones of different layers is considered. The dashed square is the region in Fig.\ \ref{fig:fig2}(b).}
\label{fig:sfig1}
\end{figure}

In Fig.\ \ref{fig:sfig1}, we show the reciprocal space of the moir\'e lattice of TBG in the extended zone scheme where equivalent Dirac points of the top (bottom) layer are shown as blue (red) dots. The form of the interlayer coupling in \eqref{eq:coupling} corresponds to scattering between Dirac cones of different layers in momentum space with the smallest momentum transfer. Higher-order Fourier component lead to longer-range ``hopping" terms in momentum space but are exponentially suppressed \cite{LopesDosSantos2012a}.

\section{TWO-BAND LATTICE MODEL} \label{app:2band}

\begin{figure}
\centering
\includegraphics[width=\linewidth]{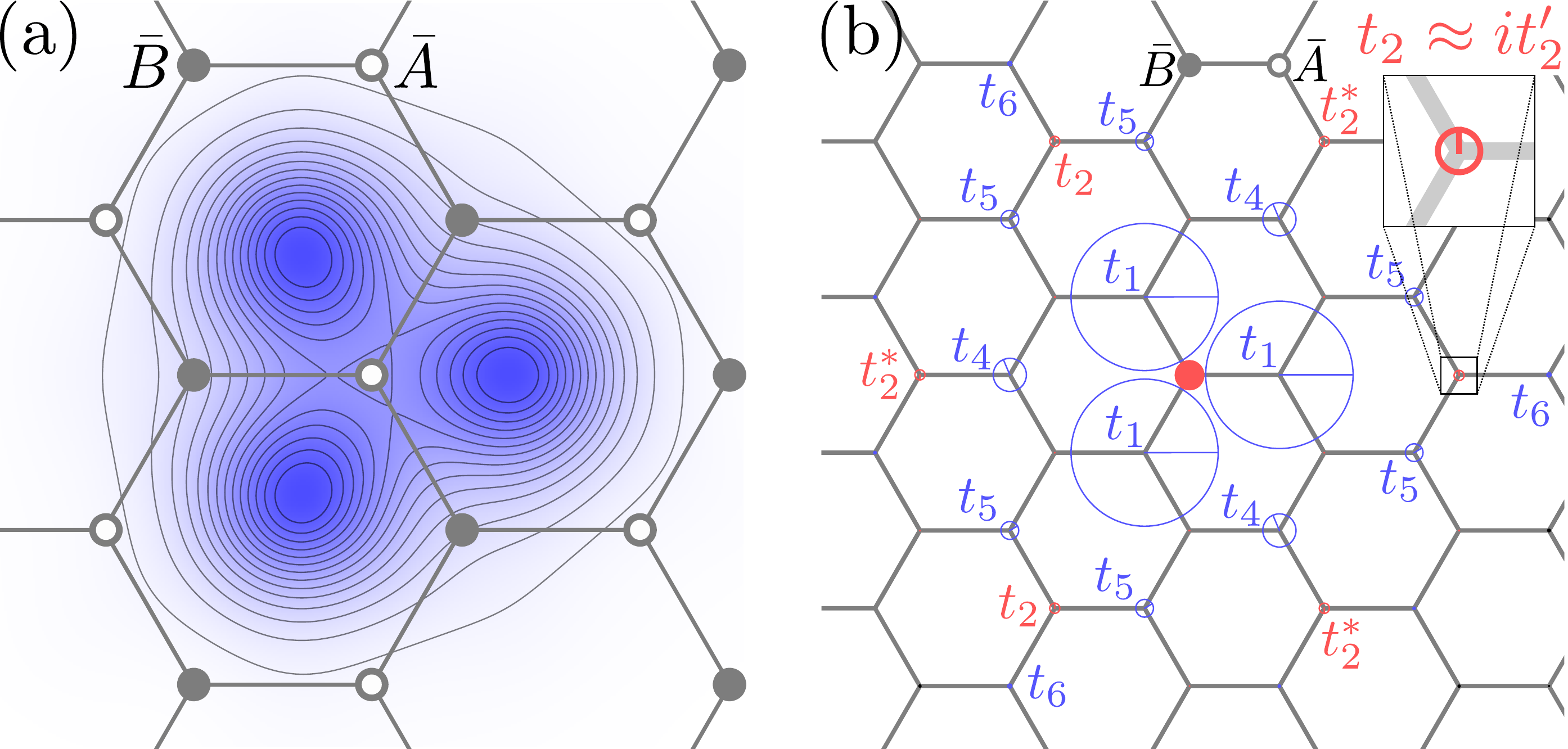}
\caption{(a) Density $|\langle \bm r | \bm R, \bar A \rangle|^2$ of the Wannier orbital centered at an $\bar A = AB$ site of the effective honeycomb lattice for $\theta = 2^\circ$. Consistent with previous studies, electrons are localized in the $AA$ regions and the orbital resembles a fidget spinner \cite{Li2010,TramblyDeLaissardiere2010,Po2018}. (b) Hopping amplitudes $t_{\sigma\bar B}(\bm R,\bm R_0)$ for $\theta=2^\circ$ from $\bar B$ at site $\bm R_0$ (red dot) to $\sigma=\bar A$ (blue) and $\sigma=\bar B$ (red) at sites $\bm R$. The circle radius corresponds to the modulus and the line indicates the phase, where we take $t_1>0$.}
\label{fig:sfig2}
\end{figure}

In this section, we follow Ref.\ \onlinecite{Koshino2018} and construct a Wannier basis from the two low-energy bands for a single valley and spin. We use a slightly different \emph{ansatz} based on their results:
\begin{align}
\hspace{-.16cm} \left| \bm R, \bar A \right> & = \frac{1}{\sqrt{N}} \sum_{\bm k} e^{-i \bm k \cdot \bm R} \frac{1}{\sqrt{2}} \left( \left| \Psi_{1\bm k} \right> + \left| \Psi_{2\bm k} \right> \right), \\
\hspace{-.16cm} \left| \bm R, \bar B \right> & = \frac{1}{\sqrt{N}} \sum_{\bm k} e^{-i \bm k \cdot \bm R} \frac{1}{\sqrt{2}} e^{i\phi_{1\bm k}} \left( \left| \Psi_{1\bm k} \right> - \left| \Psi_{2\bm k} \right> \right),
\end{align}
where $\bm R = n_1 \bm L_1 + n_2 \bm L_2$ is a moir\'e lattice vector, $N$ is the number of moir\'e cells, $\bm k$ runs over all momenta in the first MBZ, and $\left| \Psi_{m\bm k} \right>$ are the Bloch states \eqref{eq:bloch} obtained from the continuum model. Unlike Ref.\ \onlinecite{Koshino2018}, we take a gauge where the $B_t$ component of $\left|\Psi_{m\bm k}\right>$ is real at the center of the $BA$ region of the moir\'e lattice and $\phi_{m\bm k}$ is taken such that the $B_b$ component of $e^{i\phi_{m\bm k}} \left|\Psi_{m\bm k}\right>$ is real at the center of the $AB$ region. This is consistent with the symmetry analysis of Refs.\ \onlinecite{Yuan2018,Koshino2018}. In this way, the different Bloch states in $\left| \bm R, 1 \right>$ interfere constructively at the $BA$ region, and we expect it to be localized around $\bm R + \bm r_{\textrm{BA}}$. Similarly, $\left| \bm R, 2 \right>$ is expected to be localized around $\bm R + \bm r_{\textrm{AB}}$. We do not further optimize these Wannier states, so our Wannier basis is not maximally localized. However, we find that our \emph{ansatz} is well-localized above the magic angle, as shown in Fig.\ \ref{fig:sfig2}(a), and thus works well for intermediate twists ($2^\circ < \theta < 10^\circ$), which is our regime of interest.

The hopping amplitudes between the Wannier orbitals are then obtained as follows:
\begin{align} \label{eq:hopping}
& t_{\sigma \sigma'}(\bm R, \bm R') = \langle \bm R, \sigma | \hat H | \bm R', \sigma' \rangle \\
& = \frac{1}{N} \sum_{\bm k,m} e^{i \bm k \cdot \left( \bm R - \bm R' \right)} U_{\sigma'm}(\bm k) U_{m\sigma}^*(\bm k) E_{m\bm k},
\end{align}
where $\sigma,\sigma'= \bar A, \bar B$, and we used $\langle \Psi_{m\bm k} | \hat H | \Psi_{m'\bm k'} \rangle = E_{m \bm k} \delta_{mm'} \delta_{\bm k \bm k'}$, and
\begin{equation}
U(\bm k) = \frac{1}{\sqrt{2}} \begin{pmatrix} 1 & 1 \\ e^{i\phi_{1\bm k}} & -e^{i\phi_{1\bm k}} \end{pmatrix}.
\end{equation}

In the Wannier basis, the effective Hamiltonian for the two low-energy bands can then be written as
\begin{equation} \label{eq:latticemodel2}
\hat H_{\textrm{eff}} = \sum_{nn'} \sum_{\sigma\sigma'}  t_{nn'}^{\sigma \sigma'} \hat c_{n\sigma}^\dag \hat c_{n'\sigma'},
\end{equation}
where $n,m$ run over the sites of the dual honeycomb lattice with sublattices $\bar A = AB$ and $\bar B = BA$, see Fig.\ \ref{fig:sfig2}(a), on which the Wannier orbitals are centered. We also defined fermion operators $\hat c_{n\sigma}$ ($\hat c_{n\sigma}^\dag$) which destroy (create) particles in the Wannier state $\left| \bm R_n , \sigma \right>$ and hopping matrices $t_{nn'}^{\sigma \sigma'} = t_{\sigma\sigma'}(\bm R_n, \bm R_{n'})$. The hopping amplitudes are illustrated in Fig.\ \ref{fig:sfig2}(b) for $\theta=2^\circ$. Note that \eqref{eq:latticemodel2} lacks time-reversal symmetry within a single valley because the hopping amplitudes are complex in general. The hopping amplitudes for the other valley are obtained through complex conjugation, restoring time-reversal symmetry when both valleys are accounted for.

The bulk spectrum of \eqref{eq:latticemodel2}, taking into account only the largest hoppings: $t_1$, $t_2'$, $t_4$, $t_5$, and $t_6$, which are defined in Fig.\ \ref{fig:sfig2}(b), is given by
\begin{equation}
\begin{aligned}
& E_\pm(\bm k) = t_2' g(\bm k) \\
& \quad \pm \left| t_1 f_1(\bm k) + t_4 f_4(\bm k) + t_5 f_5(\bm k) + t_6 f_6(\bm k) \right|,
\end{aligned}
\end{equation}
where
\begin{align}
g(\bm k) & = 2 ( \sin \left[ \bm k \cdot \left( 2 \bm L_2 - \bm L_1 \right) \right] \\
& \quad + \sin \left[ \bm k \cdot \left( 2 \bm L_1 - \bm L_2 \right) \right] - \sin \left[ \bm k \cdot \left( \bm L_1 + \bm L_2 \right) \right] ), \nonumber \\
f_1(\bm k) & = 1 + e^{i \bm k \cdot \bm L_1} + e^{i \bm k \cdot \bm L_2}, \\
f_4(\bm k) & = e^{i \bm k \cdot \left( \bm L_1 + \bm L_2 \right)} + 2 \cos \left[ \bm k \cdot \left( \bm L_1 - \bm L_2 \right) \right], \\
f_5(\bm k) & = e^{2i \bm k \cdot \bm L_1} + e^{2i \bm k \cdot \bm L_2} + e^{-i \bm k \cdot \bm L_1} + e^{-i \bm k \cdot \bm L_2} \\
& \quad + e^{i \bm k \cdot \left( 2 \bm L_1 - \bm L_2 \right)} + e^{i \bm k \cdot \left( 2 \bm L_2 - \bm L_1 \right)}, \nonumber \\
f_6(\bm k) & = e^{-i \bm k \cdot \left( \bm L_1 + \bm L_2 \right)} + e^{i \bm k \cdot \left( 3 \bm L_1 - \bm L_2 \right)} \\
& \quad + e^{i \bm k \cdot \left( 3 \bm L_2 - \bm L_1 \right)}, \nonumber
\end{align}
with $\bm L_{1,2} = L_m ( \sqrt{3}/2, \pm 1/2 )$. The energy bands of the two-band lattice model for $\theta=2^\circ$ are shown in Figs.\ \ref{fig:sfig3}(a) and (c) where we compare it with the energy bands calculated with the continuum model.

The origin of the transverse deflection of the reflected current can be traced back to the imaginary intrasublattice hopping $t_2\approx i t_2'$ \cite{Yuan2018,Koshino2018}. In $\bm k$ space, this term becomes $\propto \tau g(\bm k) \sigma_0$ where $\tau=\pm$ corresponds to the valley and $g(\bm k)$ is an odd function with $\left. g(\bm k) \right|_{\bar M} = -4 \sqrt{3} k_x L_m + \mathcal O(k^3L_m^3)$. This pushes the VHS from $\bar M$ along the $\bar \Gamma\bar M$ line \cite{Yuan2019}. Hence, the transverse velocity near $\bar M$ is asymmetric, giving rise to transverse valley currents.

\subsection*{Two-band versus multiband model}

\begin{figure}
\centering
\includegraphics[width=.9\linewidth]{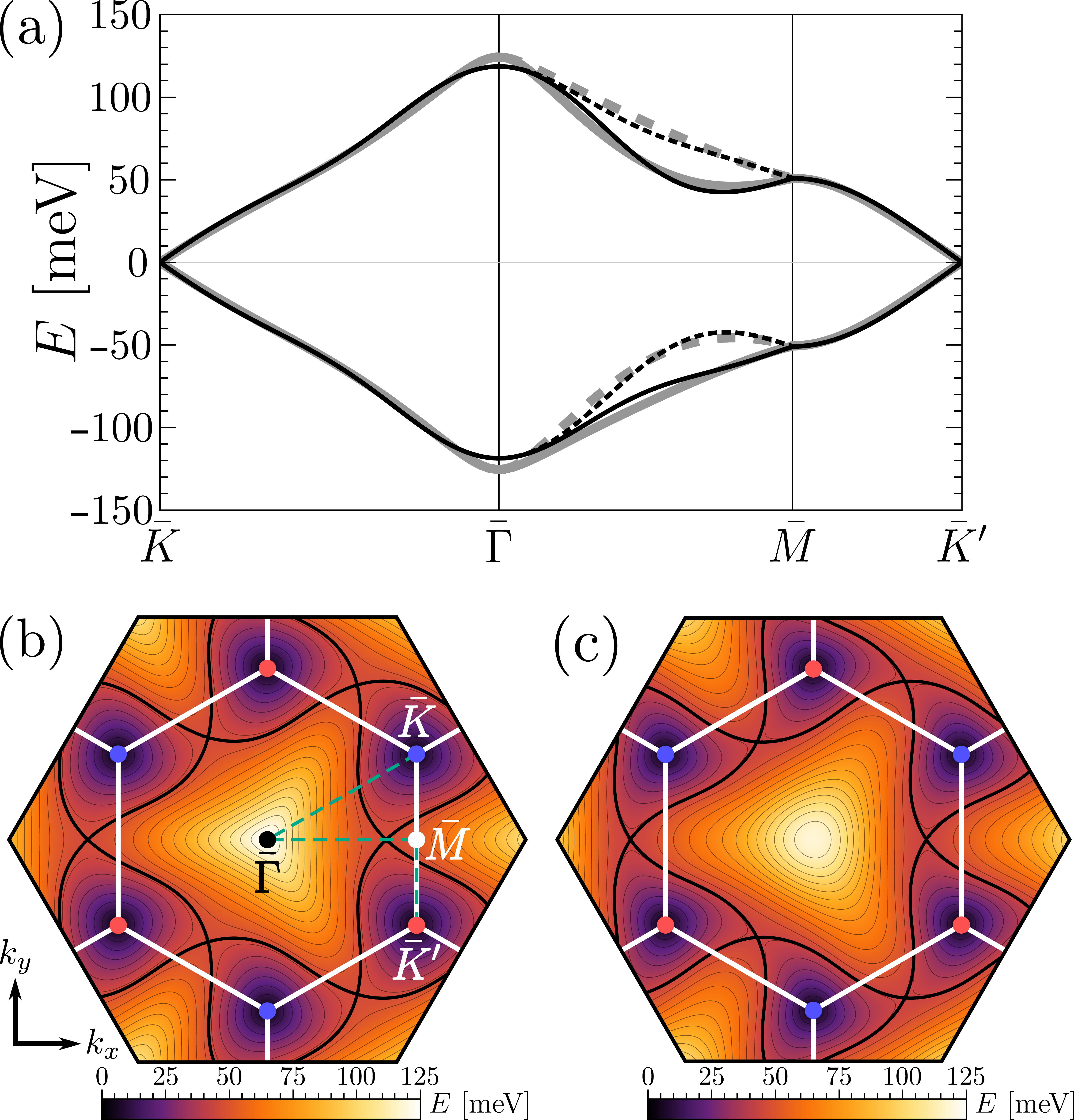}
\caption{Low-energy bands for $\theta = 2^\circ$. (a) Bands calculated with the continuum model (thick) and the lattice model (thin) for valley $K$ (solid) and $K'$ (dashed) along the path shown in (b) as the green dashed line. The parameters for the continuum model are taken from Ref.\ \onlinecite{Koshino2018} and the hoppings of the lattice model are given in Table \ref{tab:hopping}. (b) Lowest conduction band for $K$ in the continuum model. (c) Same as (b) but calculated with the two-band model. The inner white hexagon is the MBZ and the thick contour is the VHS. Red (blue) dots correspond to Dirac points of the bottom (top) layer.}
\label{fig:sfig3}
\end{figure}

The method we have used to construct the minimal two-band model introduces nonzero hoppings that break the emergent $C_2T$ symmetry of TBG in the limit of small twist angles which locally protects the Dirac points and hence the spectrum is generically gapped \cite{Po2018,Zou2018}. In our case, such symmetry-breaking terms are not included, as incidentally they are very small. However, the main purpose of the model is to reproduce the energy bands since our proposal relies on semiclassics, which it accomplishes very well as can be seen in Fig.\ \ref{fig:sfig3}. By including more bands, one can account for all symmetries without fine tuning \cite{Po2019}. We believe the most important difference relating to our results, between such a model and the two-band model we employ, is given by the chirality of the Dirac points of the two layers in the same valley. In the continuum model the chirality is the same, while in the two-band model it is opposite. This affects transport at junctions that are sharp on the moir\'e scale as this would lead to interminivalley scattering at low energies  which is sensitive to the relative chirality. However, the main results and focus of our work concerns junctions that are smooth on the moir\'e scale, in which case the chirality of the minivalleys is less important.

\section{TWO-TERMINAL SETUP} \label{app:2terminal}

\subsection{Minivalley filter}

\begin{figure}
\centering
\includegraphics[width=\linewidth]{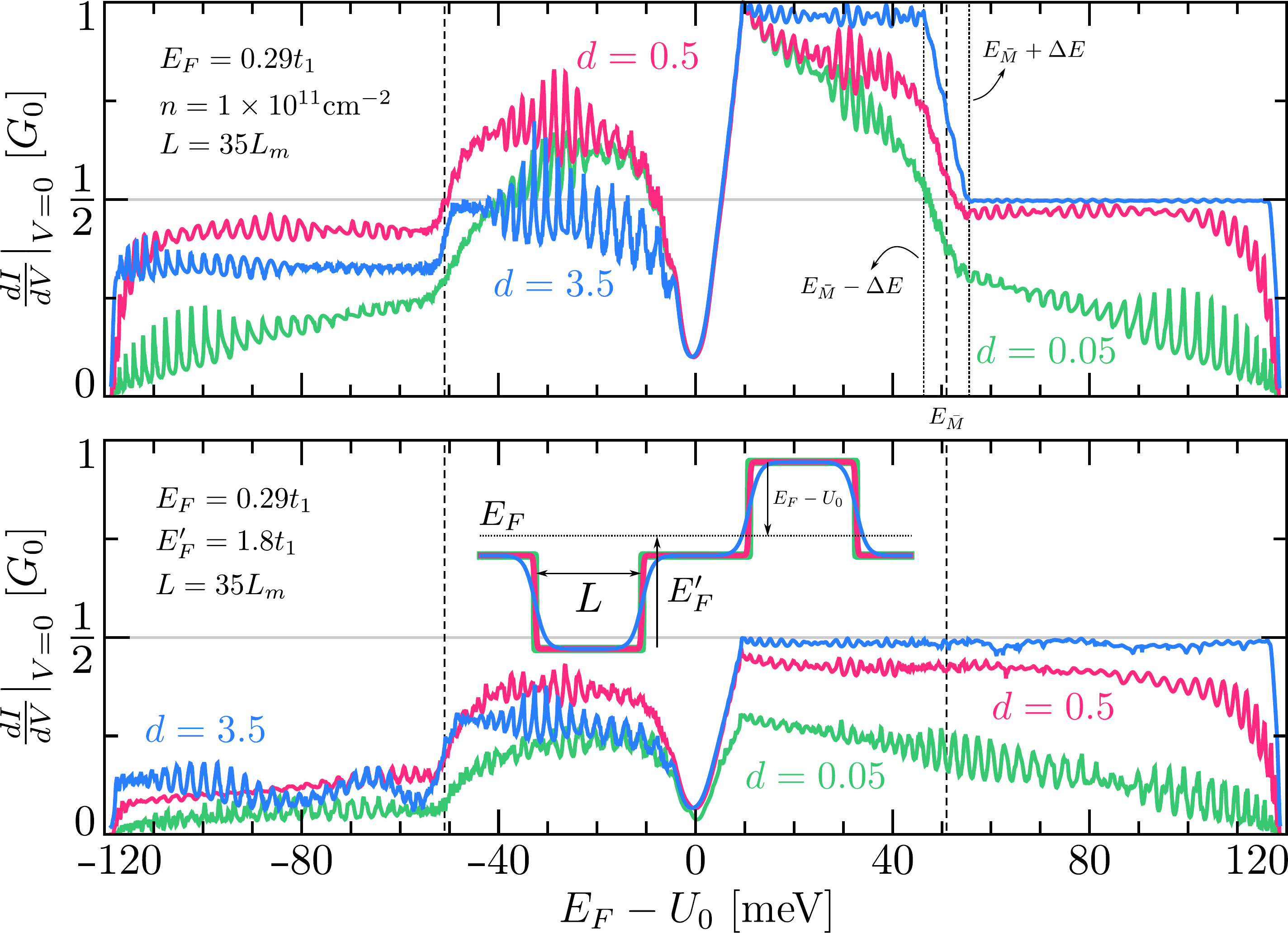}
\caption{(top) Normalized conductance for a single barrier for valley $K$ as a function $E_F-U_0$. Dashed vertical lines correspond to $\pm E_{\bar M}$. This is the same plot as the top panel of Fig.\ \ref{fig:fig3} without Gaussian averaging to smooth out oscillations. (bottom) Normalized conductance for a double barrier, see inset, for valley $K$ as a function of the Fermi energy in the second barrier, where the Fermi energy in the first barrier is constant and given by $E_F'\approx59.8$~meV. The inset show the potential profile for $d=0.05,0.5,0.35$ which determines the variation of the potential on the moir\'e scale.}
\label{fig:sfig4}
\end{figure}

The minivalley filter (i.e.\ valley splitter), illustrated in Fig.\ \ref{fig:fig1}(c), is realized for a slowly-varying potential (on the moir\'e scale) when the height (depth) of the barrier exceeds the energy at $\bar M$, $E_F-U_0>E_{\bar M}$. In this case, electrons from one minivalley propagate freely above the barrier while electrons from the other minivalley are fully reflected via the semiclassical mechanism explained in the main text. Hence, the blue line in the right part of the top panel of Fig.\ \ref{fig:sfig4} shows an almost perfect conductance plateau at $G_0/2$. In this case all transverse modes from one minivallley are reflected. However, in the range approximately given by $E_{\bar M} - \Delta E < E_F-U_0 < E_{\bar M} + \Delta E$ with $\Delta E =  \left( \partial E/\partial k_\parallel \right)_{\bar M} E_F/v_\theta$, where $v_\theta$ is the Fermi velocity of the twisted bilayer, the number of transverse modes that are reflected gradually diminishes and the conductance increases linearly from $G_0/2$ to $G_0$. When $U_0$ increases further for $E_F-U_0>0$, the conductance is close to $G_0$ with small oscillations, until the barrier height becomes comparable to $E_F$.

In the bipolar regime ($E_F-U_0<0$) one might naively also expect a minivalley filter for $E_F-U_0<-E_{\bar M}$ (for the other minivalley) since electrons in the barrier which were perfectly reflected via the semiclassical mechanism are now fully transmitted and vice versa. However, in order to get inside the barrier electrons now need to pass an $n$-$p$-junction at the left side of the barrier and then leave the barrier through the $p$-$n$ junction at the right side. Due to Klein tunneling the $n$-$p$ junction is mostly transparent when $E_F$ is sufficiently small \cite{Cheianov2006} and we still observe a conductance reasonably close to $G_0/2$ for smooth barriers (blue and red lines) as well as a high minivalley polarization of the current.
Interestingly, both the conductance plateau and minivalley polarization in the bipolar region holds better not for the smoothest potential ($d=3.5$, blue) but for the intermediate one ($d=0.5$, red). To understand this, we need to consider another length scale present in our system. In order for electrons to have a well-defined minivalley that corresponds well to the layers (left panel of Fig.\ \ref{fig:fig2}), $E_F$ needs to be close to the Dirac point. At the same time, the validity of the semiclassical mechanism requires a smooth potential on the scale given by the VHS energy $\lambda_{\textrm{VHS}} \sim \hbar v_\theta/E_{\textrm{VHS}} \sim L_m$. However, electrons at the $n$-$p$-junction have a much larger wavelength $\lambda_F \sim \hbar v_\theta/E_F$. If the length scale of the barrier potential is large compared to $\lambda_F$, tunneling through the junction is suppressed \cite{Cheianov2006}. Hence the reduced conductance and minivalley polarization for $d=3.5$ (blue lines). On the other hand, if the potential scale is small compared to $\lambda_F$ (still large compared to $\lambda_{\textrm{VHS}}$), the barrier at the $n$-$p$ junction becomes effectively sharp. The total transmission probability for a sharp junction is independent of $E_F$ and turns out to be close to unity. This is because the transmission probability reaches unity at normal incidence due to Klein tunneling and for sharp junctions the width of this maximum effectively spans the entire range of transverse modes. In the absence of interminivalley processes, scattering at an $n$-$p$ junction for $E_F$ close to the Dirac point is described by the Hamiltonian
\begin{equation}
H= \sigma_x p_x + \sigma_y p_y + U(x), \qquad H \Psi = \varepsilon \Psi,
\end{equation}
where we put $\hbar v_F=1$ and which corresponds to a generic crossing in the presence of the potential $U(x)$. Here, we take $|U(x)| \ll |\varepsilon|$ for $x < 0$ and $U(x)>0$ with $U(x) \gg |\varepsilon|$ for $x>0$. For concreteness, we take $\varepsilon>0$ from now on. In this case, the right-moving scattering state for $x>0$ can be approximately written as
\begin{equation}
\Psi_> = \bigg( \begin{array}{c} 1 \\ 1 \end{array} \bigg) \exp{\left[ - i \int_0^x dx' \, U(x') \right]},
\end{equation}
while at $x<0$, the wave function is given by
\begin{equation}
\Psi_< = u_+ e^{i \left( p_x x + p_y y \right)} + u_- e^{i \left( -p_x x + p_y y \right)},
\end{equation}
where we take $p_x>0$ and the spinors $u_+,u_-$ may be chosen as
\begin{equation}
u_+ = a \begin{pmatrix} \varepsilon \\ p_x + i p_y \end{pmatrix}, \quad u_- = b \begin{pmatrix} \varepsilon \\ -p_x + i p_y \end{pmatrix}.
\end{equation}
The coefficients $a$ and $b$ are found from matching the two solutions at $x=0$:
\begin{equation}
a = \frac{p_x-ip_y+\varepsilon}{2p_x \varepsilon}, \qquad b = \frac{p_x+ip_y-\varepsilon}{2p_x \varepsilon}.
\end{equation}
Current conservation is assured as 
\begin{align}
\Psi_<^\dag \sigma_x \Psi_< & = u_+ \sigma_x u_+ + u_- \sigma_x u_- \\
& = 2p_x \varepsilon \left( |a|^2 - |b|^2 \right) \\
& = 2 = \Psi_>^\dag \sigma_x \Psi_>.
\end{align}
We thus find the reflection probability
\begin{equation}
R = \frac{|b|^2}{|a|^2} = \frac{\varepsilon - p_x}{\varepsilon + p_x},
\end{equation}
where both $\varepsilon>0$ and $p_x>0$. For $\varepsilon<0$, the sign of $p_x$ is also reversed as electrons in the valence band propagate opposite to their momentum and hence the result holds in general. The normalized conductance is then given by
\begin{equation}
\frac{G}{G_0} = \frac{1}{2E_F} \int_{-E_F}^{E_F} dp_y \left( 1 - R \right) = 4 - \pi \approx 0.858,
\end{equation}
which is valid for a single $n$-$p$ junction. For two $n$-$p$ junctions on either side of the barrier, we find
\begin{equation}
\frac{G_2}{G_0} = \frac{1}{2E_F} \int_{-E_F}^{E_F} dp_y \, \frac{1-R}{1+R} = \frac{\pi}{4} \approx 0.785,
\end{equation}
where we neglected interference effects. We believe that this is the reason for the red curve ($d=0.5$) for $E_F-U_0<-E_{\bar M}$ showing better filtering than the blue curve ($d=3.5$) in the bipolar regime. On the other hand, if in the unipolar filtering regime ($E_F-U_0>E_{\bar M}$) the potential barrier is not sufficiently smooth with respect to $\lambda_F$, electrons can be reflected at the ``sharp" edge of the barrier. That is why the minivalley filter for $d=0.5$ is similarly imperfect for both the bipolar $E_F-U_0<-E_{\bar M}$ and unipolar $E_F-U_0>E_{\bar M}$ regime.

\begin{figure}
\centering
\includegraphics[width=\linewidth]{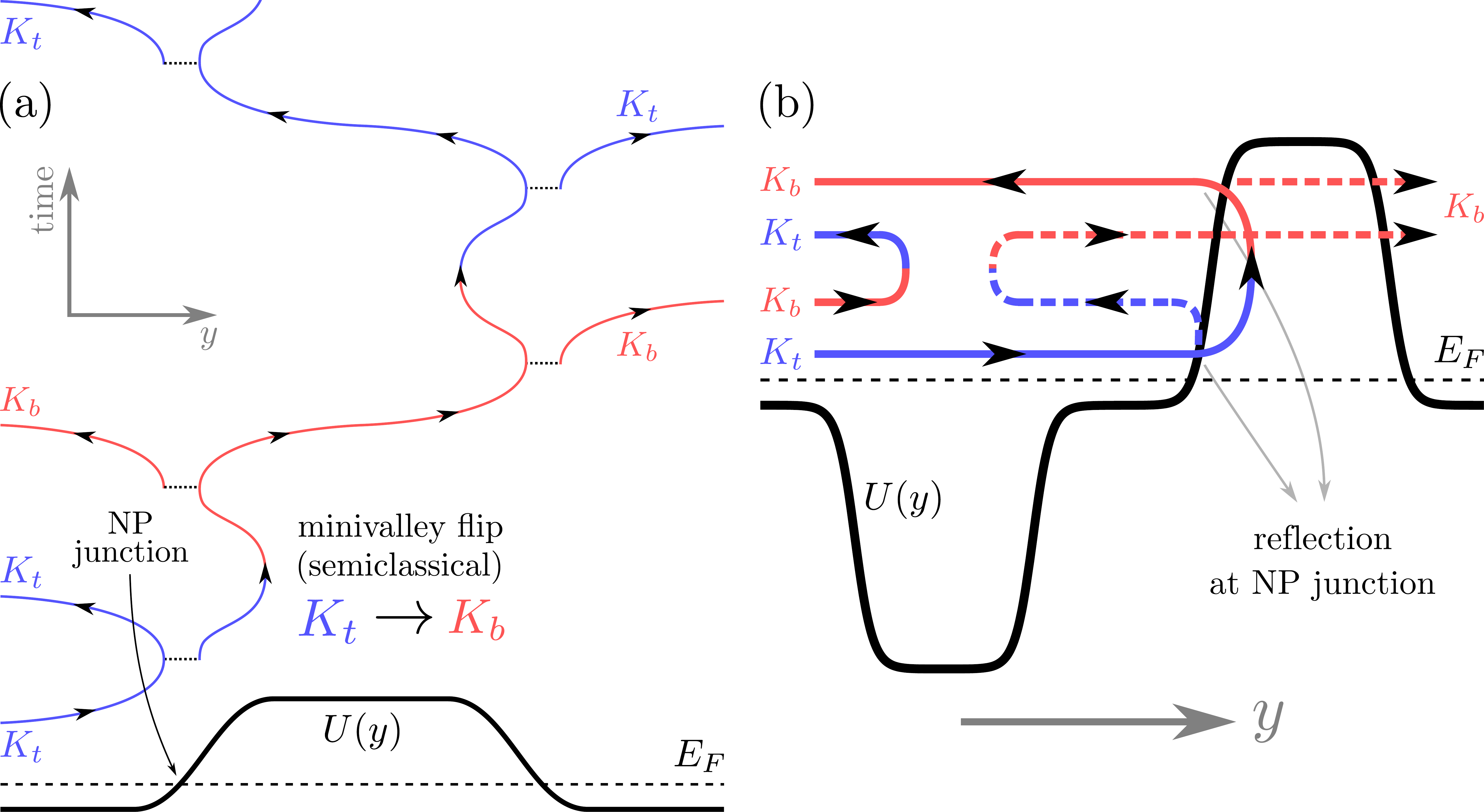}
\caption{(a) Electron trajectories in the bipolar regime with $E_F-U_0<-E_{\bar M}$. (b) Trajectories for the valley valve. Here, the first barrier is in the unipolar regime ($E_F-U_0>E_{\bar M}$) such that it is transparent (opaque) for right-moving (left-moving) modes at minivalley $K_t$ and vice versa for $K_b$, while the second is in the bipolar regime ($E_F-U_0<-E_{\bar M}$). The valley valve does not block all current as a small fraction of electrons are reflected at the $n$-$p$ junctions without changing minivalley, shown as the dashed lines (for a single reflection).}
\label{fig:sfig5}
\end{figure}

The conductance oscillations in the bipolar regime appear due to the interference with electrons bouncing inside the barrier \cite{Silvestrov2007}. The interference mechanism in the bipolar filtering regime is illustrated in Fig.\ \ref{fig:sfig5}(a). Here, we sketch the trajectories of an electron, which in the absence of the $n$-$p$ junction (e.g.\ for $E_F<0$) would be reflected due to the semiclassical mechanism shown in Fig.\ \ref{fig:fig1}(c). The trajectory is first split into transmitted and reflected parts at the left $n$-$p$ junction. The transmitted part is then reflected inside the barrier which changes the minivalley (blue to red). This trajectory goes back to the left junction and is split again. Note that the transmitted (red) wave does not interfere with the previously reflected (blue) wave as they correspond to different minivalleys. Electrons bouncing inside the barrier start emitting interfering waves only after a full period consisting of two reflections that flip the minivalley and four reflections at two $n$-$p$ junctions. This may be the reason why the oscillations are stronger in the bipolar Dirac regime ($-E_{\bar M} < E_F - U_0 < 0$) than in the bipolar filtering regime ($E_F -U_0 < -E_{\bar M}$) because of the smaller length of interfering trajectories in the former. 

Finally, if $|E_F-U_0|$ exceeds the bandwidth of the two lowest bands, the conductance drops to zero as there are no modes in the minigap. For the smooth potential the conductance drops to zero in almost the same way as the drop from $G_0$ to $G_0/2$ at $E_F-U_0 > E_{\bar M}$ where one of the minivalleys is blocked. This is because the conductance drop can also be viewed as the barrier becoming opaque for both minivalleys.

\subsection{Valley valve}

Ideally the barrier only transmits electrons from one minivalley/layer for a given valley, such that one might design a valley valve \cite{Rycerz2007,Akhmerov2008} with two barriers in series. This setup is illustrated by the inset of the bottom panel of Fig.\ \ref{fig:sfig4}. When the two barriers are gated similarly, one minivalley can go through. For opposite polarities, one would hope that all current is blocked. Indeed, electrons that pass the first barrier, and therefore belong to the minivalley for which the second barrier is opaque, are reflected by the semiclassical mechanism in the second barrier which flips the minivalley. The reflected electrons now belong to the correct minivalley to pass the first barrier from the other side and hence all electrons are reflected back. This process is illustrated in Fig.\ \ref{fig:sfig5}(b). 

In reality, not all current is blocked due to the presence of $n$-$p$ junctions on the left and right side of one of the barriers. Scattering at these junctions, which are smooth on the moir\'e scale, conserves the minivalley as the momentum $k_\theta \propto L_m^{-1}$ cannot be provided. Therefore, after the electrons are reflected semiclassicaly in the second barrier, they can reflect back at the $n$-$p$ junction without changing the minivalley and pass the barrier. Such processes spoil the valley valve but are suppressed when the barrier is still sharp on the scale of $\lambda_F$ as we discussed in the previous section. Hence, there is a trade-off between smooth junctions for which the minivalley filtering works better and less smooth junctions for which there is less reflection at the $n$-$p$ junctions.

Theoretically, there is no small parameter that ensures that the reflection at a sharp $n$-$p$ junction (relative to $\lambda_F$) is strongly suppressed. Nevertheless, we find that the valley valve works reasonably well as shown in the bottom panel of Fig.\ \ref{fig:sfig4}. This is again because of the fact that Dirac electrons at normal incidence and effectively many other transverse modes are passing through without reflection (Klein tunneling). In the figure, we see that when the barriers have the same polarities, the conductance is approximately given by $G_0/2$ for the slow-varying potential ($d=3.5$) and $E_F-U_0>E_{\bar M}$ since only one of the minivalleys is transmitted. For opposite polarities, the conductance is suppressed for $E_F-U_0<-E_{\bar M}$.

\section{SIX-TERMINAL SETUP} \label{app:6terminal}

\begin{figure}
\includegraphics[width=\linewidth]{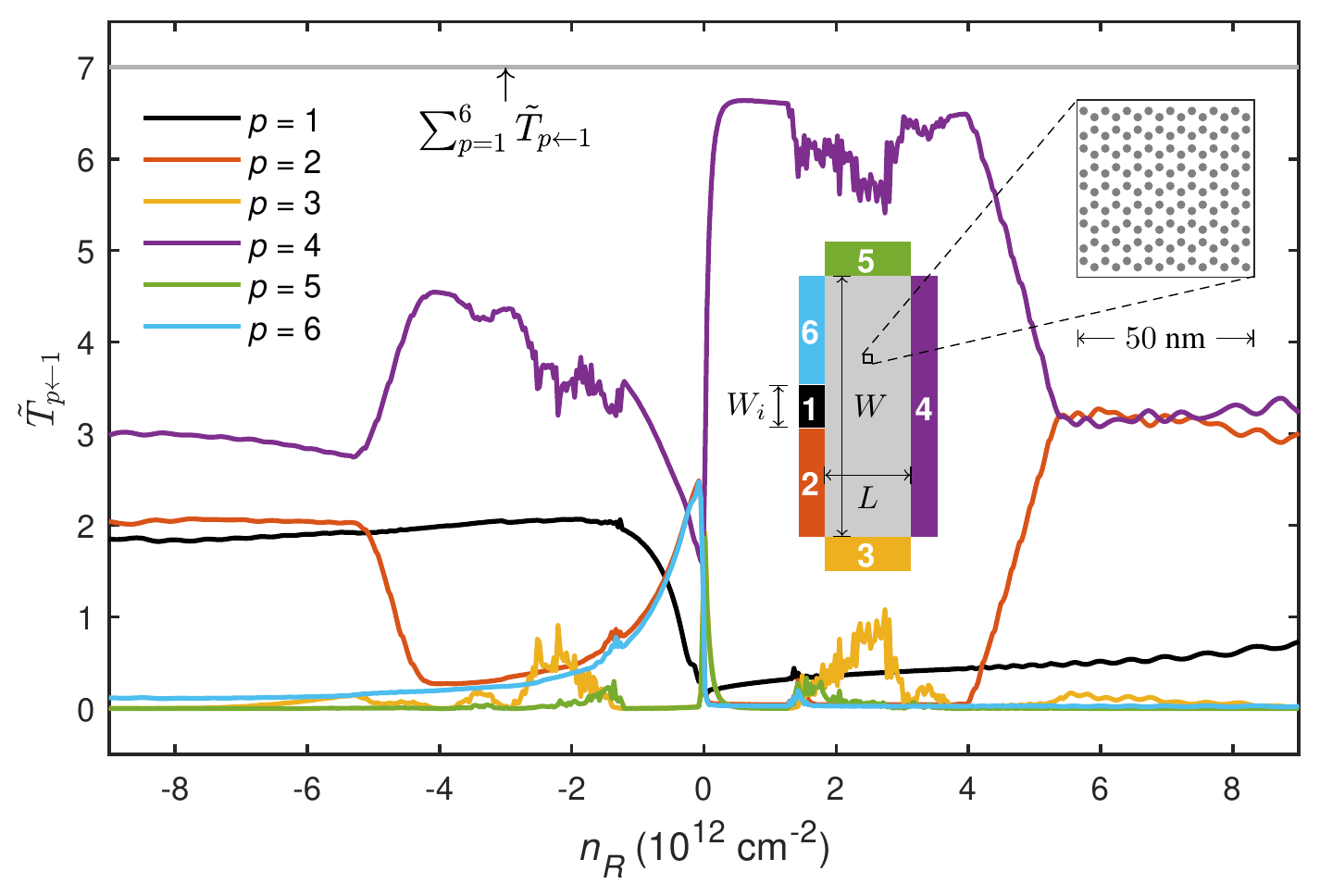}
\caption{Transmission functions of the six-terminal setup considered in Fig.\ \ref{fig:fig4} of the main text.}
\label{fig:figS_T}
\end{figure}

In the multiterminal case, we considered a scattering region of length $L=1$~$\mu$m and width $W=3$~$\mu$m attached to six leads in the presence of a smooth junction (see Fig.\ \ref{fig:fig4}). The scattering region and the leads are described by the effective lattice model for twist angle $\theta=2^\circ$. See the insets of Fig.\ \ref{fig:figS_T}, where the transmission functions $\tilde{T}_{p\leftarrow 1}$ from lead $1$ to lead $p=1,2,\cdots,6$ are shown. These six transmission functions sum to the number of modes $M_1$ of the incoming lead $1$, as required by the sum rule \cite{Datta1995}
\begin{equation}
\sum_{p=1}^6 \tilde{T}_{p\leftarrow q} = M_q,
\end{equation}
where $M_q$ is the number of incoming modes in lead $q$. The horizontal gray line in Fig.\ \ref{fig:figS_T} shows that the above sum rule is fulfilled with $M_1=7$. The normalized transmission functions that are shown in Fig.\ \ref{fig:fig4}(c) as well as in Fig.\ \ref{fig:figS_normalizedT} are defined by
\begin{equation}
T_{p\leftarrow 1} = \frac{\tilde{T}_{p\leftarrow 1}}{M_1} \ . 
\end{equation}

\subsection{Width of the injection lead}

\begin{figure}
\includegraphics[width=\linewidth]{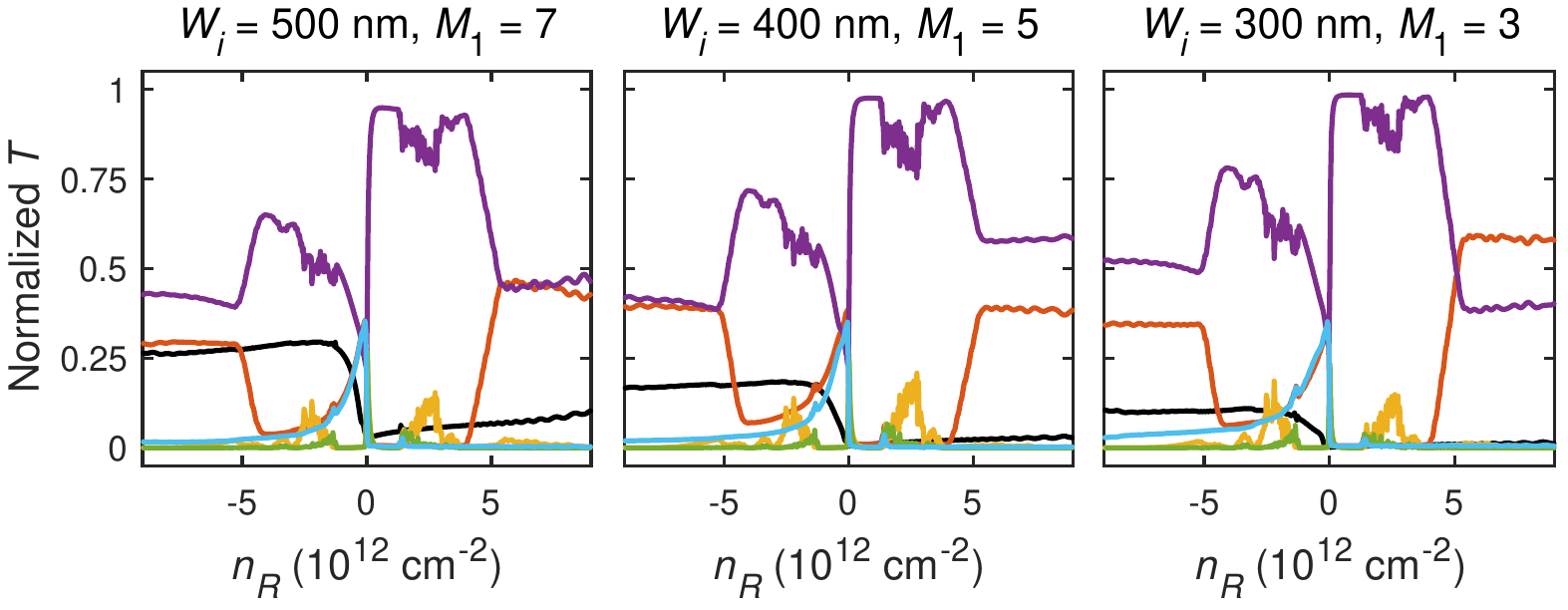}
\caption{Normalized transmission functions of the six-terminal setup for different widths $W_i$ of the injection lead as indicated. The coloring is the same as in Fig.\ \ref{fig:figS_T}.}
\label{fig:figS_normalizedT}
\end{figure}

Figure \ref{fig:figS_normalizedT} shows the normalized transmission functions for various widths of the injection lead 1, where we take $n_L=10^{11}$~cm$^{-2}$ which is the same as in Fig.\ \ref{fig:figS_T} as well as in Fig.\ \ref{fig:fig4}(c). In the figure, we see that the expected values $\tilde{T}_{2\leftarrow 1}\approx\tilde{T}_{4\leftarrow 1}\approx 0.5$ are best realized with a wider injection lead. Overall, the transmission functions shown in Fig.\ \ref{fig:figS_normalizedT} exhibit similar behaviors for different widths of the injection lead.
\begin{figure*}
\centering %
\includegraphics[scale=0.43]{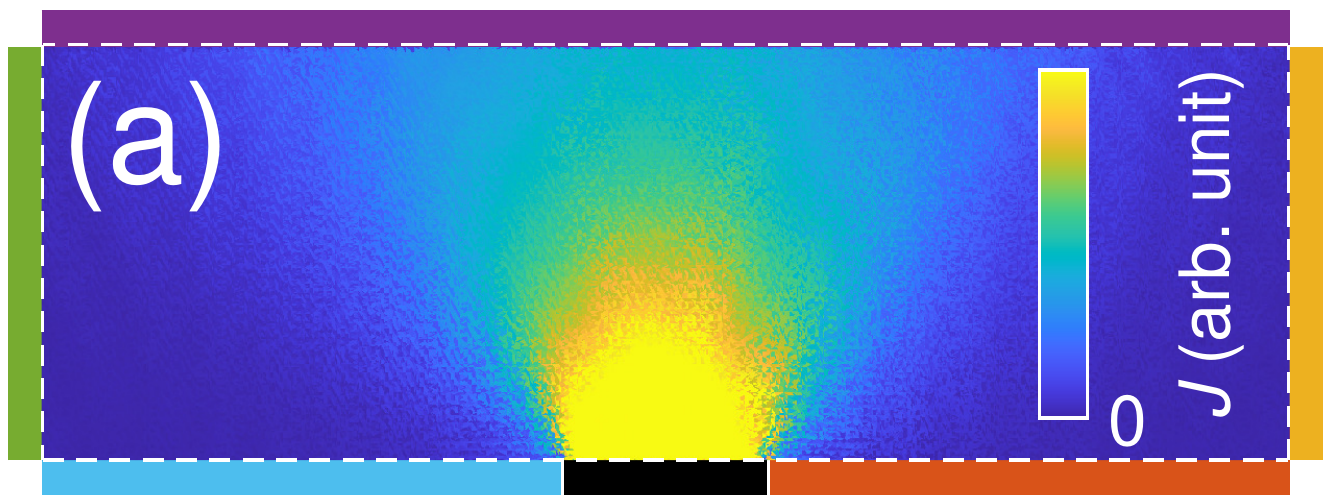}
\includegraphics[scale=0.43]{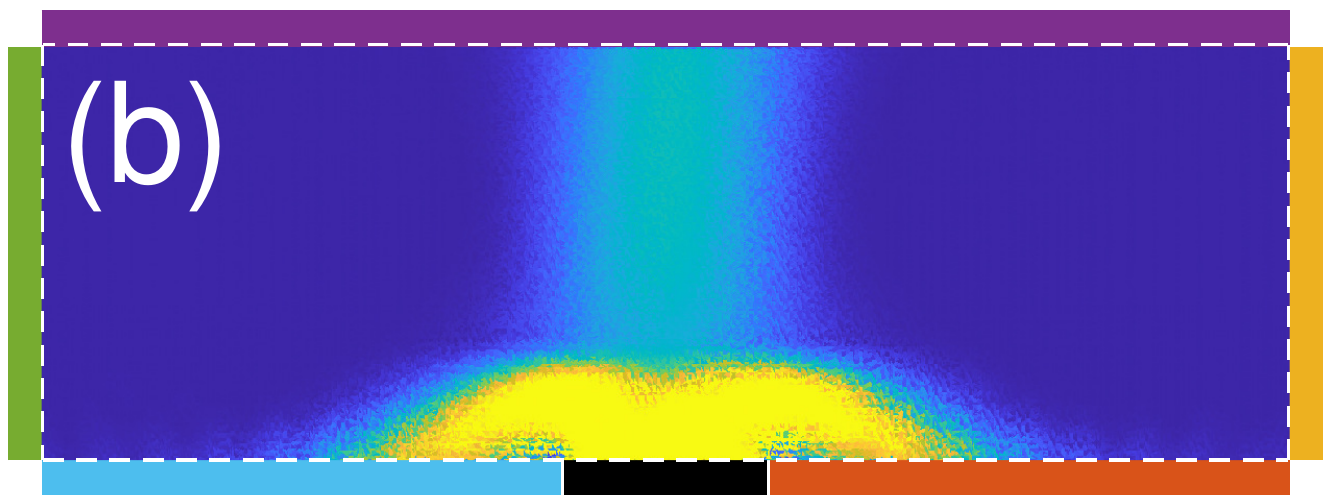}
\includegraphics[scale=0.43]{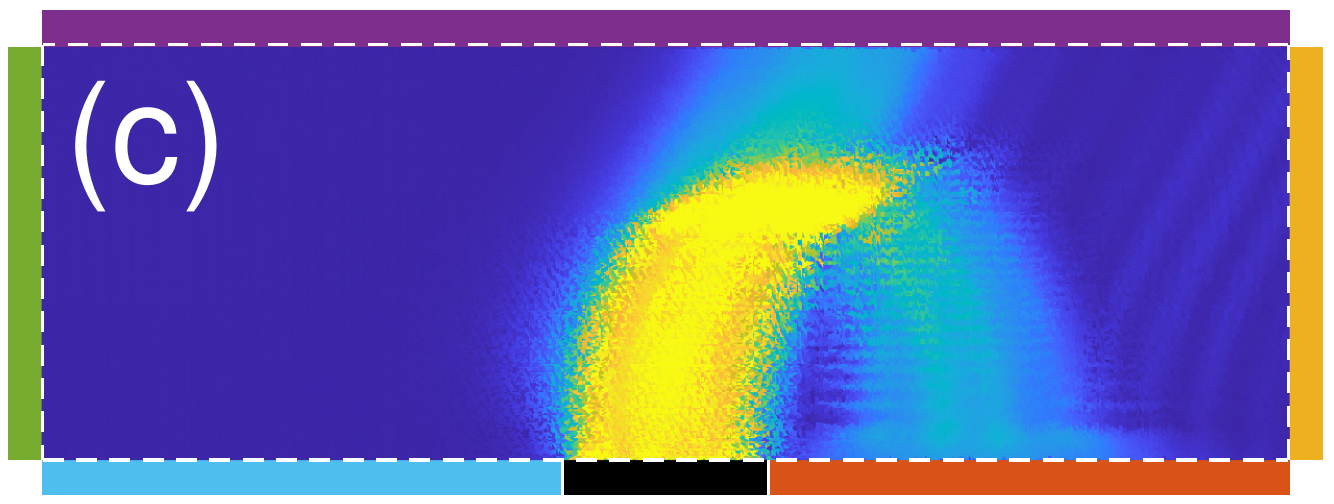} \medskip
\includegraphics[scale=0.43]{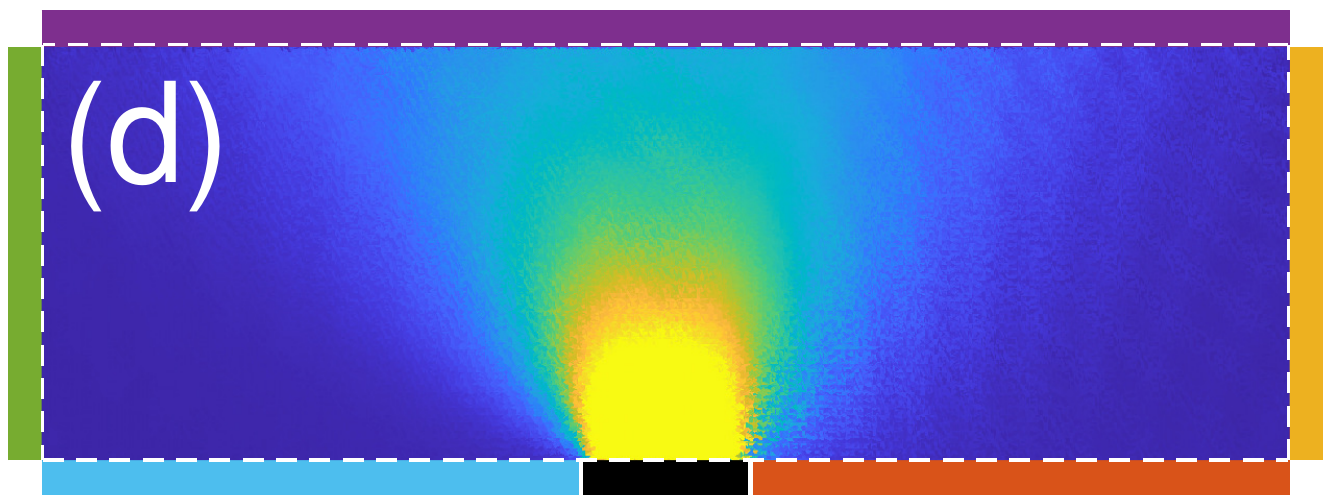}
\includegraphics[scale=0.43]{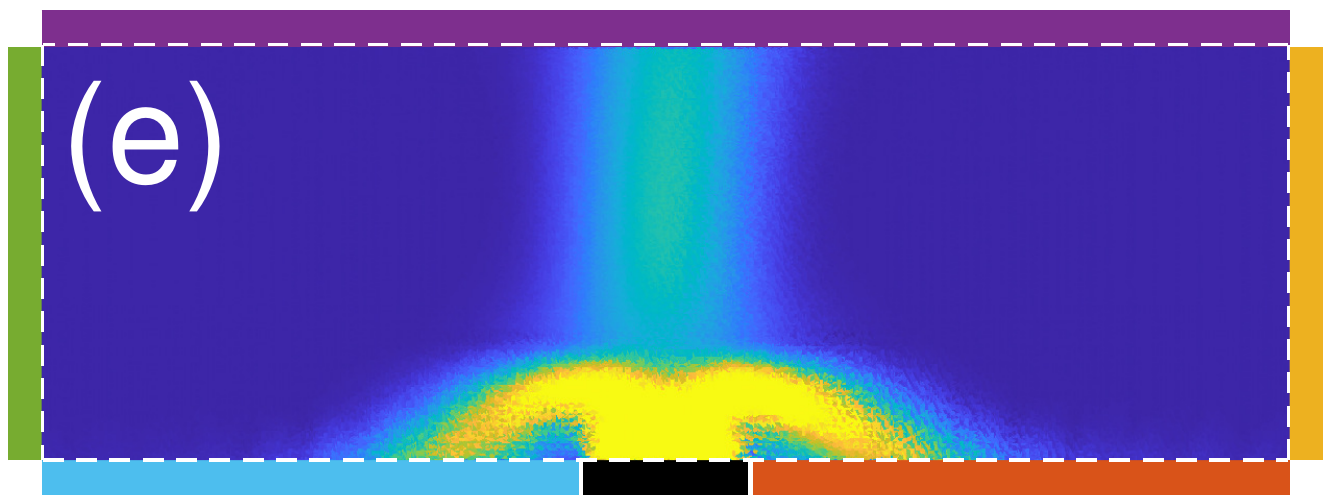}
\includegraphics[scale=0.43]{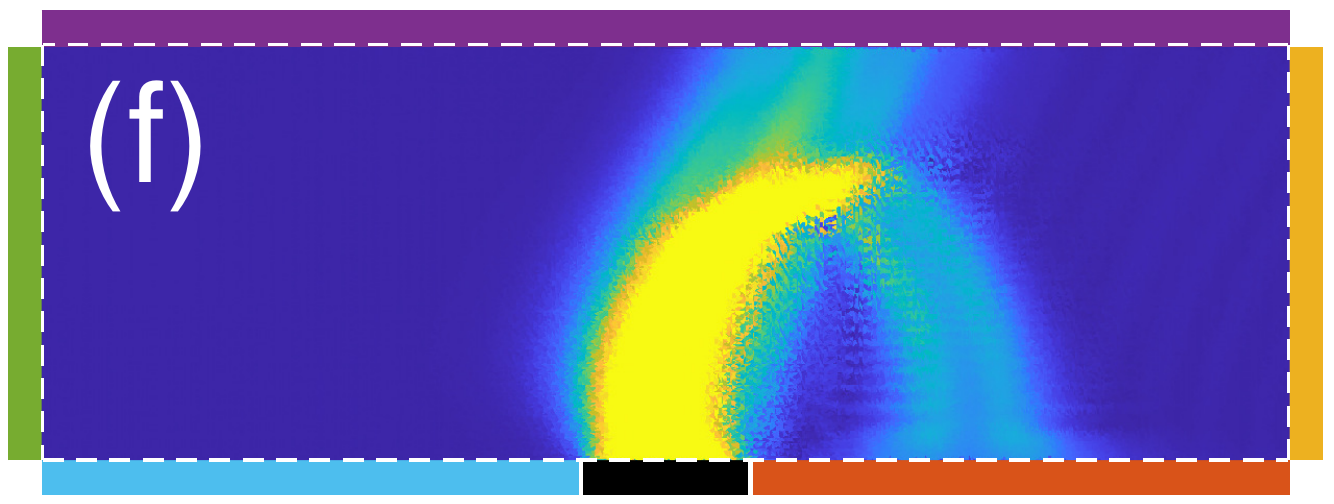} \medskip
\includegraphics[scale=0.43]{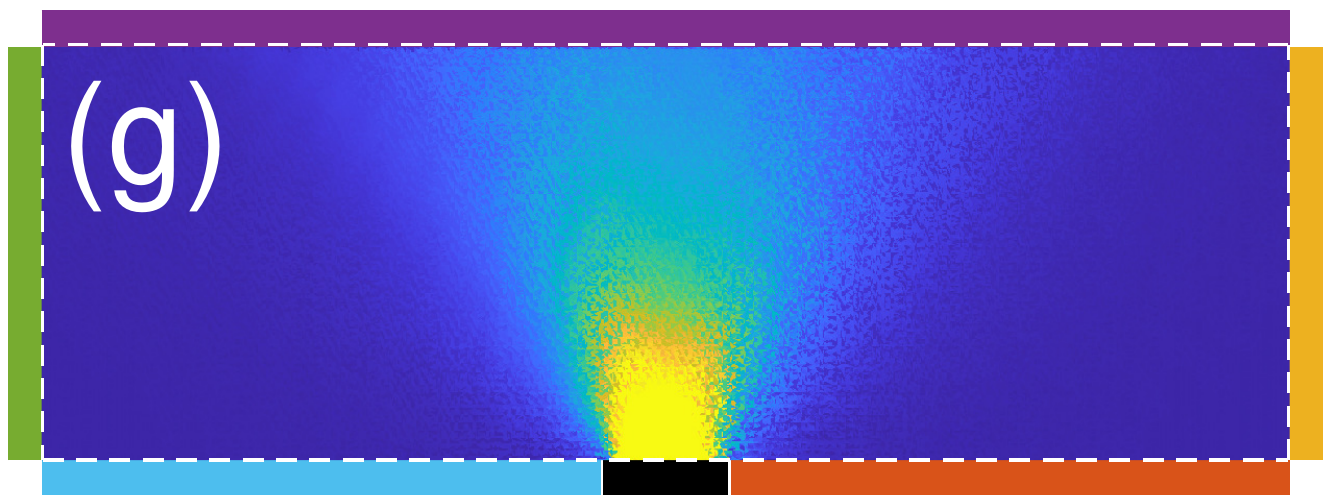}
\includegraphics[scale=0.43]{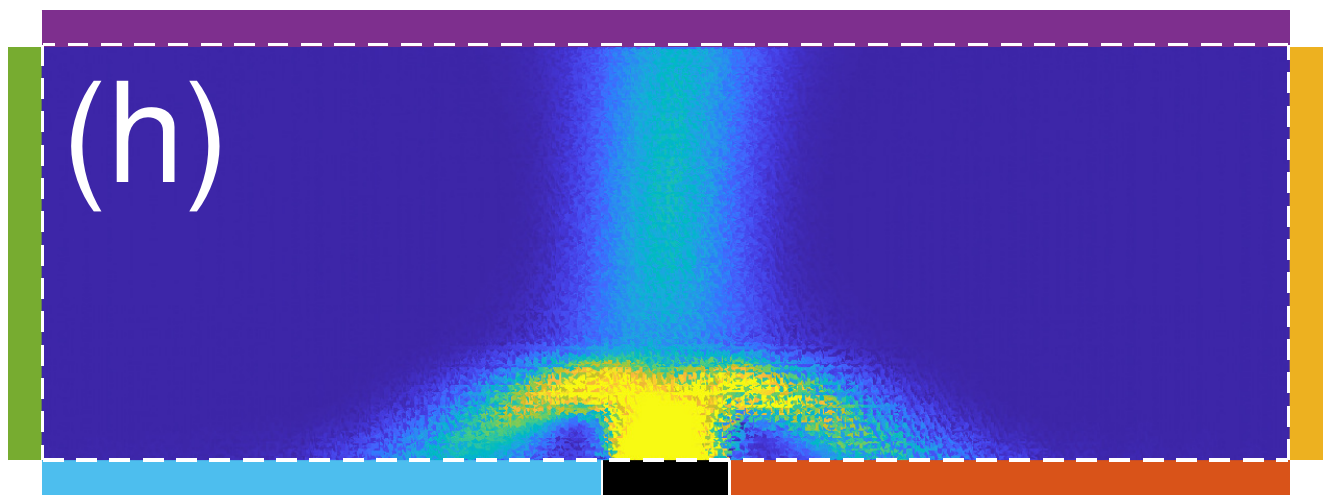}
\includegraphics[scale=0.43]{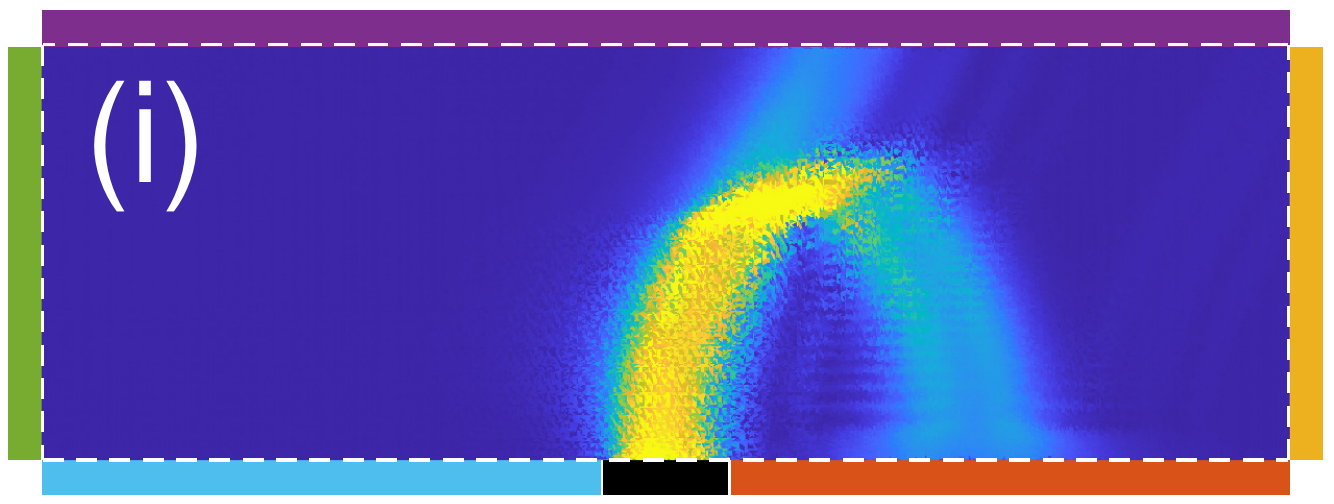}
\caption{Local current densities of the six-terminal setup, subject to $n_L=1\times 10^{11}$~cm$^{-2}$ and $n_R=2.3\times 10^{11}$~cm$^{-2}$ for the left column of panels (a), (d), and (g), $n_R=-2.3\times 10^{11}$~cm$^{-2}$ for the middle column of panels (b), (e), and (h), and $n_R=6.7\times 10^{11}$~cm$^{-2}$ for the right column of panels (c), (f), and (i). The width of the injection lead is $W_i=500$~nm for (a)--(c), $W_i=400$~nm for (d)--(f), and $W_i=300$~nm for (g)--(i). All panels share the same color bar displayed in (a).}
\label{fig:figS_J}
\end{figure*}

\subsection{Local current density}

In the main text, we have shown a few current density profiles to illustrate the valley-selective transverse current deflection. Here, we show more profiles for different density configurations ($n_L=10^{11}$~cm$^{-2}$ remains fixed) and injection lead widths but with the other parameters the same as before ($L=1~\mu$m, $W=3~\mu$m, $\theta=2^\circ$). Note that we only consider the $K$ valley here, while the $K'$ valley is obtained from $K$ by time reversal.

In Fig.\ \ref{fig:figS_J}, the first row (a)--(c), the second row (d)--(f), and the third row (g)--(i) correspond to $W_i=500$, $400$, and $300$~nm, respectively. Despite the decreasing incoming number of modes (see Fig.\ \ref{fig:figS_normalizedT}), the current flow behaves similarly for different $W_i$. The first column of Fig.\ \ref{fig:figS_J} [panels (a), (d), and (g)] considers the case $(n_L,n_R)= (1.0,2.3) \times 10^{11}$~cm$^{-2}$, so that the TBG system is in the low-energy unipolar state, showing typical single-layer graphene $nn'$ behavior. The second column of Fig.\ \ref{fig:figS_J} [panels (b), (e), and (h)] considers the case $(n_L,n_R)= (1.0,-2.3)\times 10^{11}$~cm$^{-2}$, corresponding to the low-energy bipolar state. In this case, the beam structure arises from the Klein collimation due to the smooth $n$-$p$ junction, which is also typical behavior of single-layer graphene \cite{Cheianov2006}. Finally, the third column of Fig.\ \ref{fig:figS_J} [panels (c), (f), and (i)] considers $(n_L,n_R)= (1.0,6.7)\times 10^{11}$~cm$^{-2}$, revealing the valley-selective current deflection, insensitive to the width of the injection lead. Note that Figs.\ \ref{fig:figS_J}(b), (c), and (i) are identical to Figs.\ \ref{fig:fig4}(d), (e), and (f), respectively.

\subsection{Classical trajectories}

The classical trajectories shown in Fig.\ \ref{fig:fig4}(a), are calculated with the Hamilton equations,
\begin{equation}
\dot{\bm k} = -\nabla_{\bm r} U, \qquad \dot{\bm r} = \nabla_{\bm k} E(\bm k).
\end{equation}
where we replaced the classical Hamiltonian $H_{\text{classical}} \rightarrow E(\bm k) + U(y)$, where $E(\bm k)$ is the energy of the first conduction band of TBG, and the potential profile $U(y)$ is determined by the relation between the electron density $n$ and $E$ shown in Fig.\ \ref{fig:sfig5}(b) in order to obtain a linear density profile
\begin{equation}
n(y) = 
\begin{cases} n_L & y < 0, \\
n_L + \left( n_R - n_L \right) y / y_0 & 0 < y < y_0, \\
n_R & y > y_0.
\end{cases}
\end{equation}
Even though the two-band model with the hoppings we consider has inversion symmetry, the Berry curvature does not strictly vanish as time-reversal symmetry is effectively broken in each valley separately. However, since the hoppings that break time reversal in a single valley, i.e.\ $t_2'$, $\textrm{Im}~t_4$, $\textrm{Im}~t_5$, and $\textrm{Im}~t_6$ are small, we have neglected the anomalous velocity. This assumption is warranted by the good comparison between the semiclassical trajectories and the transport calculation, see Fig.\ \ref{fig:fig4}(f).

\bibliography{references.bib}

\end{document}